\documentclass{scrartcl}
\usepackage{amsmath}
\usepackage{amsfonts}
\usepackage{amssymb}
\usepackage{amsthm}

\usepackage{dsfont}
\usepackage[T1]{fontenc}
\usepackage{epsfig}
\usepackage{graphicx}
\usepackage{natbib}
\usepackage{enumerate} 
\usepackage{multirow}
\usepackage{xcolor}
\usepackage{enumitem}

\usepackage{thmtools}

\usepackage{comment}

\usepackage[textsize=footnotesize,  linecolor=red!80!white,  backgroundcolor=yellow!20!white,bordercolor=red!80!white]{todonotes}

\newcommand{\N}{\mathbb{N}}
\newcommand{\R}{\mathbb{R}}

\newcommand{\Rd}{\mathbb{R}^d}

\newcommand{\D}{{\mathcal{D}}}

\newcommand{\PROB}{{\mathbf P}}
\renewcommand{\P}{{\cal P}}
\newcommand{\F}{{\cal F}}

\renewcommand{\bf}{\normalfont \bfseries}
\renewcommand{\it}{\normalfont \itshape}

\usepackage{geometry}
\usepackage[utf8]{inputenc}

\usepackage{dashrule}
\usepackage[labelfont=bf]{caption} 
\usepackage{cleveref}
\usepackage{tikz}
\usepackage{pgfplots}
\usetikzlibrary{calc, arrows}

\usepackage{subfig}

\usepackage{float}

\allowdisplaybreaks
\begin{document}
\renewcommand{\thefootnote}{\fnsymbol{footnote}}

\begin{center}

  {\LARGE \bf
    Uncertainty Quantification in Case of Imperfect Models:
    A Review
  }
\footnote{
Running title: {\it Uncertainty quantification: a review}}
\vspace{0.5cm}

Sebastian Kersting\footnote{
Corresponding author. Tel: +49-6151-16-23374, Fax:
+49-6151-16-23381} and 
Michael Kohler

{\it 
Fachbereich Mathematik, Technische Universit\"at Darmstadt,
Schlossgartenstr. 7, 64289 Darmstadt, Germany,
email: kersting@mathematik.tu-darmstadt.de, kohler@mathematik.tu-darmstadt.de}

\end{center}
\vspace{0.5cm}

\begin{center}
December 11, 2020
\end{center}
\vspace{0.5cm}

\noindent
    {\bf Abstract}\\
    Uncertainty quantification of complex technical systems
    is often based on a computer model of the system.
    As all models such a computer model is always wrong 
    in the sense that it does not describe the reality
    perfectly. The purpose of this article is to give
    a review of techniques which use observed values
    of the technical systems in order to take into account
    the inadequacy of a computer model in uncertainty
    quantification. The techniques reviewed in this article
    are illustrated and compared by applying them to 
    applications in mechanical engineering.
    
\vspace*{0.2cm}


\vspace*{0.2cm}


\section{Introduction}
\label{se:introduction}

Uncertainty quantification is a major research field of statistical methods with applications in engineering sciences. 
Methods of uncertainty quantification are often used to analyse experiments with technical systems. 
These experiments can be described by $ \Rd \times \R $-valued (random) variables $ (X,Y) $, where $ X $ describes input parameters of the experiments and $ Y $ describes the outcome of the experiment.
E.g. if one considers drop tests of a spring damper as described in Subsection \ref{se:MAFDS} the experimental outcome depends on the drop height, i.e. in this case the input dimension $ d $ is equal to $ 1 $ and the measured outcome is the maximal relative compression.

Usually to conduct real experiments is expensive and time consuming. To circumvent this problem computer models that simulate the experiment with the technical system are playing crucial role. 
\cite{SaWiNo2003} and \cite{FaLiSu2010} provide an overview of methods for the design and analysis of experiments conducted with these computer models.

Computer models are often computationally expensive and thus it is not possible to conduct a large quantity of computer experiments. To circumvent this problem one can use so-called surrogate models. 
There exists different methods to estimate a surrogate model, e.g. \cite{BuBo1990}, \cite{KiNa1997}. \cite{DaZh2000} used quadratic response surfaces. \cite{Hu2004}, \cite{DeLe2010}, \cite{BoDeLe2011} investigated surrogate models in context of support vector machines, \cite{PaLa2002} concentrated on neural networks, \cite{Ka2005} and \cite{Bietal2008} used kriging.
Usually these surrogate models do not take an inherent error of the computer model into account and thus are themself imperfect in case of an imperfect computer model.

There is a number of different objectives in methods of uncertainty quantification. The main objectives are computer model calibration, computer model validation and quantifying the results of an experiment.
\cite{KeOh2001} and \cite{HaSaRa2009} proposed a Bayesian method for computer model calibration and model validation based on Gaussian processes, \cite{Hietal2013} proposed a similar approach as \cite{KeOh2001}, but used the ensemble Kalman filter. \cite{Pl2017} also  introduced a Bayesian computer model calibration method similar to the approach of \cite{KeOh2001}, but their prior distribution on the bias (model discrepancy) is modeled orthogonal to the gradient of the computer model.
\cite{Baetal2007} proposed a validation method in case of time dependent systems, where they model the discrepancy between the computer experiments and the outcome of the technical system by a Gaussian process.
\cite{WaChTs2009} proposed a Bayesian method for model validation.
\cite{GuWa2018} proposed the so-called scaled Gaussian Process and use it for model calibration. They claim, that the method bridges the gap between the $ L_{2} $ calibration and the Gaussian process calibration.
\cite{Daetal2018} proposed a method for computer model calibration based on perfect and imperfect computer models.
\cite{TuWu2015} proposed the $ L_{2} $ calibration model, to identify unknown computer model parameter. They also pointed out that the Bayesian approach of \cite{KeOh2001} might fail in case of an imperfect computer model, since there exists no values of the parameters which fit the technical system perfectly.
\cite{Pl2019} proposed a frequentist method for computer model calibration by constructing confidence intervals for the model parameter.

Based on perfect computer models, \cite{DeFeKo2013} and \cite{BoFeKo2015} derived consistency and rate of convergence results for density estimators based on surrogate models, \cite{FeKoKr2015a} and \cite{FeKoKr2015b} proposed a method for the adaptive choice of smoothing parameters for density estimators based on surrogate models and \cite{KoKr2019} proposed a method to estimate quantiles based on surrogate models.

The articles mentioned above either assume that the underlying computer model is perfect, or only deal with computer model calibration. Normally, computer models are imperfect, i.e. they do not predict the outcome of real experiments perfectly, e.g. because of a relaxation of underlying physical dependencies (typically neglecting the friction or by considering it to be constant) to reduce complexity or because of missing knowledge about the technical system.
In this article we give an overview of existing methods based on imperfect computer models with the following goals.
The first aim is to construct an improved surrogate model and based on that quantify the outcome of the experiment $ Y $ either by density or quantile estimation. The second is to quantify the model error and thus enable the comparison of different computer models. The last aim is to quantify the influence of the computer model error on the quantification of $ Y $.

\subsection{Notation}
Throughout this paper we use the following notation:
$\N$, $\R$ and $\R_+$  are the sets of positive integers, real numbers, and nonnegative real numbers, respectively.
For $z \in \R$ we denote the smallest integer greater than or equal to $z$ by $\lceil z \rceil$.
For $x \in \Rd$ we denote the $i$-th component of $x$ by $x^{(i)}$.
If $X$ is a random variable, then $\PROB_X$ is the corresponding distribution, i.e., the measure associated with the random variable.
Let $D \subseteq \R^d$ and let $f:\R^d \rightarrow \R$ be a real-valued function defined on $\R^d$. We write 
$x = \arg \min_{z \in D} f(z)$ if $\min_{z \in \D} f(z)$ exists and if $x$ satisfies
\[
x \in D \quad \mbox{and} \quad f(x) = \min_{z \in \D} f(z).
\]
If $ A $ and $ B $ are sets, with $ A \subseteq B $. then $ I_{A}\colon B \to \{0,1\} $ is the indicator function corresponding to $ A $, i.e. the function which takes on the value 1 on A and is 0 elsewhere.

\subsection{Outline}
The outline of this paper is as follows:
In Section \ref{se:technical_systems} we introduce two technical systems, which will be used in the following sections to illustrate and discuss the described methods.
In Section \ref{se:data_models} we describe typical data models and a method used to generate additional input values.
In Section \ref{se:uq_improved_surrogate_models} uncertainty quantification methods based on improved surrogate models are described and illustrated.
In Section \ref{se:computer_model_comparision} methods used to compare different computer simulations are described and illustrated.
In Section \ref{se:confidence_estimator} two methods which can be used to quantify the influence of the computer model error are described and illustrated.

\section{Technical systems}
\label{se:technical_systems}
In the following, two technical systems are introduced. The data generated by experiments with these systems will be used to illustrate the methods of uncertainty quantification.


\subsection{Modular active spring damper system}
\label{se:MAFDS}

\begin{figure}[h!]
	\centering
	\includegraphics[width=6cm]{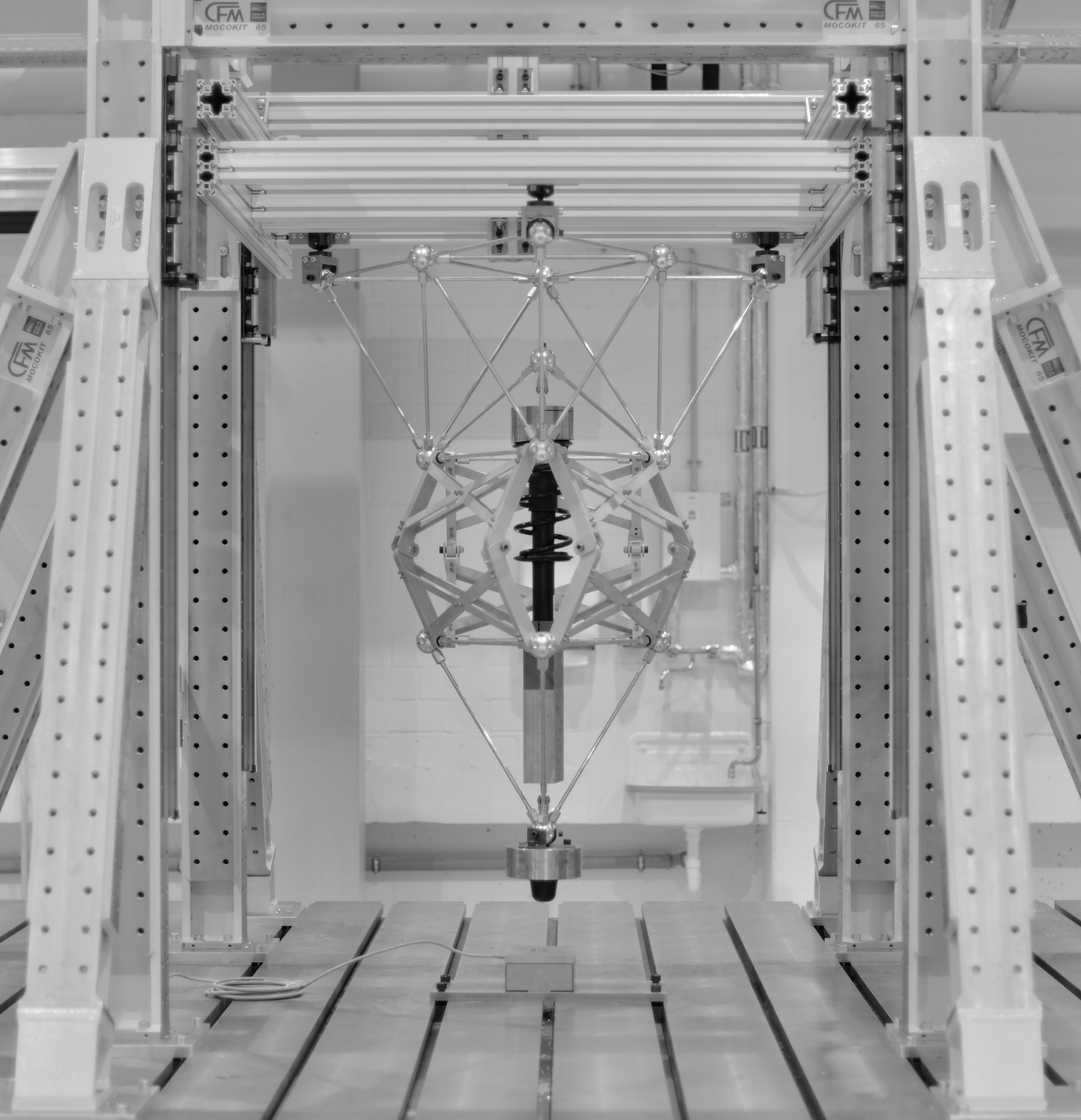}
	\caption{\label{fig:MAFDS_real} A photo of the MAFDS and its experimental test setup.}
\end{figure}

The modular active spring damper system (German acronym MAFDS) shown in  Figure \ref{fig:MAFDS_real} is a suspension system that was designed with similar specifications and requirements as an air plane landing gear, although it is not a landing gear substitute.  
It was developed in the collaborative research centre SFB 805 at the Technische Universität Darmstadt in order to investigate uncertainty in a load-bearing structural system when predicting the dynamic response. 

Its main components are an upper truss structure, a lower truss structure, guidance links that enable relative translation of the truss structures in vertical direction and a spring-damper. A detailed system description can be found in \cite{FePl2019} and \cite{MaPa2019}.
Dynamic drop tests can be carried out similar to landing gear testing: the MAFDS is lifted up and dropped by a variable drop height $h$. 
In a total of $ 100 $ experiments,   the inputs $h$ were chosen normally distributed with mean $ 0.05~\mathrm{m}$  and a standard deviation of $0.0057~\mathrm{m}$. 
As system output we regard the maximal relative compression $z_{\mathrm{r,max}}$ between the upper and the lower truss during the drop test.

Modelling of the system at hand yielded two different computer models 
to compute the maximum relative compression  $z_{\mathrm{r,max}}$ to a corresponding input value.   
With both computer models we conduct $ 500 $ simulations each and furthermore generate $ 16.000.000 $ additional input values that will be subsequently used in this paper.

\subsection{Piezo-elastic beam}
\label{se:piezo_beam}

\begin{figure}[h!]
\centering
\includegraphics[width=10cm]{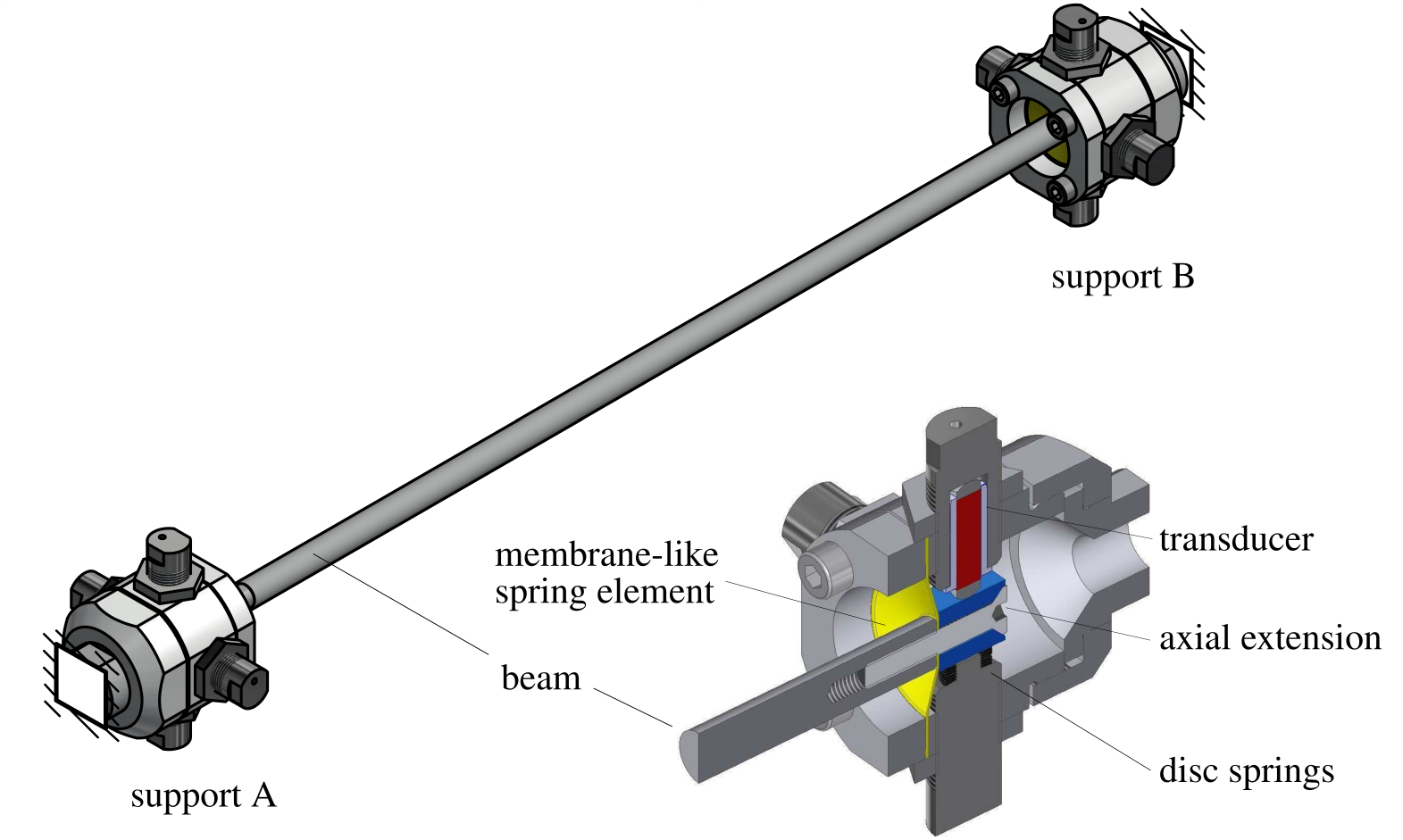}
\caption{\label{fig:piezo_beam}  A CAD model of the
lateral vibration attenuation system with
piezo-elastic supports and a sectional view of one of the
piezo-elastic supports, cf. \cite{Lietal2017}.}
\end{figure}

The second technical system under consideration is a lateral vibration attenuation system with piezo-elastic supports shown in Figure \ref{fig:piezo_beam}.
This system consists of a beam with circular cross-section embedded in two piezo-elastic supports A and B where support A is used for lateral beam vibration excitation and support B is used for lateral beam vibration attenuation, as proposed in \cite{Goetal2016} and \cite{Go2019}. 
Both supports are located at opposite beam ends and each consists of one elastic membrane-like spring element made of spring steel, two piezoelectric stack transducers arranged orthogonally to each other and mechanically pre-stressed with disc springs as well as the relatively stiff axial extension made of hardened steel that connects the piezoelectric transducers with the beam. For vibration attenuation in support B, optimally tuned electrical shunt circuits are connected to the piezoelectric transducers.

As system output serves the maximum lateral vibration amplitude $y$ in an experiment with this system. 
Five quantities of the membrane in the attenuation system vary during the manufacturing of the piezo-elastic supports and influence the maximal vibration amplitude: the lateral stiffness in $y$-direction $k_{lat,y}$ and in $z$-direction  $k_{lat,z}$, the rotatory stiffness in $y$-direction $k_{rot,y}$ and in $z$-direction   $k_{rot,z}$, and the height of the membrane $h_x$.
Ten measurements of corresponding parameters together with the experimental outcome $y$ for ten built systems are available, which are listed in Table \ref{table:data_piezo_beam}.
\begin{table}[h!]
\begin{center}
\begin{tabular}{|c|c|c|c|c|c|c|c|c|c|c|}
\hline
& 1 & 2 & 3 & 4 & 5 & 6 & 7 & 8 & 9 & 10\\
\hline
$ k_{rot,y} \times 10^{2}$ & 1.31   & 1.34   & 1.31   & 1.23   & 1.14   & 1.29   & 1.35   & 1.28   & 1.04   & 1.20   \\ 
$ k_{rot,z} \times 10^{2}$  & 1.31   & 1.28   & 1.43   & 1.25   & 1.30   & 1.34   & 1.22   & 1.16   & 1.18   & 1.11   \\
$ k_{lat,y} \times 10^{7} $ & 3.27   & 3.28   & 3.35   & 3.29   & 3.22   & 3.26   & 3.19   & 3.54   & 3.21   & 3.42   \\
$ k_{lat,z} \times 10^7 $   & 3.07   & 3.22   & 3.29   & 3.25   & 3.30   & 3.18   & 3.16   & 3.51   & 3.37   & 3.44   \\
$ h_x      \times 10^{-4}$  & 6.79   & 6.77   & 6.82   & 6.80   & 6.79   & 6.76   & 6.81   & 6.74   & 6.68   & 6.84   \\
$y \times 10^{1}$           & 1.45   & 1.42   & 1.44   & 1.42   & 1.43   & 1.35   & 1.47   & 1.32   & 1.31   & 1.63   \\
\hline
\end{tabular}
\caption{Measurement data for ten built systems. The values of $ k_{rot,y} $ and $ k_{rot,z} $ are given in $[Nm / \operatorname{rad}] $, the values of $ k_{lat,y} $ and $ k_{lat,z} $ are given in $[ N / m ]$, the values of $ h_{x} $ are given in $ [m] $ and the values of the maximal vibration amplitude $ y $ are given in $[\frac{m}{s^2}/V]$.}
\label{table:data_piezo_beam}
\end{center} 
\end{table} 

A computer model is available with which we can compute the maximal vibration amplitude $y$ to the corresponding input values of  $k_{rot,y}$, $k_{rot,z}$, $ k_{lat,y}$, $k_{lat,z}$ and $ h_x $.
Apart from the input values displayed in Table \ref{table:data_piezo_beam} additional input values are generated. 
Therefore, we assume that the input values are independent multivariate normally distributed and we generate a sample of additional input parameters as described in Subsection \ref{se:generating_inputs}. 
For simulations of the computer model, a total of $200$ additional input values are generated. 
In addition, $ 16.000.000 $ additional input values are generated that will be subsequently used in this paper. 

\section{Data models}
\label{se:data_models}
In an application, usually one of the following two different data models are available:
\begin{enumerate}[label=(\roman*)]
	\item {\bf Full information model}:
	In this case, a computer simulation $ m \colon \Rd \to \R $ of the experiment with the technical system is available and can be evaluated at arbitrary values, an experimental data set 
	\begin{equation}
	\label{eq:freq_experimental_data}
	(X_{1},Y_{1}),\ldots,(X_{n},Y_{n})
	\end{equation}
	of size $ n \in \N $ is available and the underlying distribution $ \PROB_{X} $ of $ X $ is known thus one can generate an additional data set
	\begin{equation}
	\label{eq:freq_additional_input_data}
	X_{n+1},\ldots,X_{n+L_{n}+N_{1,n}+N_{1,n}}
	\end{equation}
	of additional $ L_{n}+N_{1,n}+N_{2,n} $ input values.
	
	\item {\bf Experimental setup model}: In this case, a computer simulation $ m \colon \Rd \to \R $ of the experiment with the technical system and the experimental data set \eqref{eq:freq_experimental_data} of size $ n \in \N $ are available. Here the underlying distribution $ \PROB_{X} $ of $ X $ is unknown. 
\end{enumerate}

In the following section we will describe a method to generate an additional sample of input values.
Thus, to simplify the presentation we will assume that the full information model is available for the description of the methods.
Note that for the MAFDS, cf. Subsection \ref{se:MAFDS}, data model (i) is available and for the piezo-beam, cf. Subsection \ref{se:piezo_beam}, data model (ii) is available.

\subsection{Generating additional input values}
\label{se:generating_inputs}
To apply methods of uncertainty quantification, often a large sample of input data is necessary. In some settings (especially if the input data is random as well as the outputs) only a small data set of input values is available. In this case it is necessary to generate more input data. But this is sometimes nearly as expensive and time consuming as producing more output data, e.g. in the application from Subsection \ref{se:piezo_beam} one has to produce a new disk spring and measure the four stiffness parameters and the height parameter. Thus generating more experimental input data is often not possible.

The randomness in input data is often induced by production margins or measurement errors and thus making a realistic assumption on the underlying class of distributions is often possible. A variety of methods for data generation in case of known distributions is given in \cite{De1986}. In the following, the in \cite{KeKo2019} proposed method to generate multivariate normally distributed random variables is described:

Assume $ X $ is a multivariate normally distributed random variable with values in $ \Rd $. Furthermore assume that an independent and identically distributed sample
\begin{equation}
X_{1},\ldots,X_{n}
\end{equation}
of $ X $ is available.

Then estimate the parameters of the distribution of $ X $ by a maximum likelihood estimate defined by 
\begin{equation}
\hat{\mu} = \frac{1}{n} \sum_{i = 1}^{n} X_{i}
\end{equation}
and
\begin{equation}
\hat{\Sigma} = \left(\frac{1}{n} \sum_{k=1}^{n} (X_{k}^{(i)} - \hat{\mu}^{(i)})(X_{k}^{(j)} - \hat{\mu}^{(j)}) \right)_{1 \leq i,j \leq d}.
\end{equation}
The estimated values are then treated as if they were the real values.
Then one can generate a sample of size $ N_{n} \in \N $ which is independent and multivariate normally distributed with mean $ \hat{\mu} $ and covariance $ \hat{\Sigma} $ by firstly generating an independent sample $ Z_{1},\ldots,Z_{N_{n}} $ of $ d $-dimensional vectors, where for each vector the components are independent and standard normally distributed, and set for every $ i = 1,\ldots,N_{n} $ 
\begin{equation}
\label{se4sub2eqTemp1}
\bar{X}_{i} = \hat{O} \hat{\Lambda}^{1/2} Z_{i} + \hat{\mu},
\end{equation}
where $ \hat{O} $ and $ \hat{\Lambda} $ are defined by the eigendecomposition
\begin{equation*}
\hat{\Sigma} = \hat{O} \hat{\Lambda} \hat{O}^T
\end{equation*}
of $\hat{\Sigma}$. Here $ \hat{\Lambda} = \operatorname{diag}(\hat{\lambda}_{1},\ldots,\hat{\lambda}_{d}) $ is a diagonal matrix consisting of eigenvalues of $ \hat{\Sigma} $ and $ \hat{O} $ is a orthogonal matrix whose columns are eigenvectors of $ \hat{\Sigma} $.

\section{Uncertainty quantification based on improved surrogate models}
\label{se:uq_improved_surrogate_models}
As discussed above, computer models are usually imperfect. In this section, a method to estimate an improved surrogate model and it's application in uncertainty quantification is described.
\subsection{Estimating an improved surrogate model}
\label{se:surrogate_model}
Computer simulations of experiments with a technical system are often complex and computationally expensive to evaluate.
But for a proper analysis of the underlying technical system it is necessary to generate a large sample of these computer experiments.
In addition, the inherent error in the computer simulation, cf. Section \ref{se:introduction}, will bias the results.
To circumvent these problems, a solution is to compute an improved surrogate model and use it instead of the computer simulation.

The first step is to generate a sample of input values evaluated with the computer simulation.
\cite{KoKr2017a} and \cite{GoKeKo2018} used a sample of the input quantity $ X $, \cite{WoStLe2017} suggested to use data generated in a specific range by Latin Hypercube sampling (see \cite{McBeCo1979}), if the input data is not random, and \cite{KeKo2019} used a sample uniformly distributed on a centered cube to construct an estimator of a computer model.

The second step is to estimate a surrogate model of the computer simulation. There exists a vast variety of methods, some already mentioned above. One possibility is to use (penalized) least squares estimates, defined by 
\begin{equation}
\label{eq:freq_surrogate}
\hat{m}_{L_{n}}(\cdot) = \arg \min_{f \in \F  } \frac{1}{L_{n}} \sum_{i = 1}^{L_{n}} | f(X_{i})  - m(X_{i})|^2 + pen_{n}^2(f),
\end{equation} 
where $ \F $ is a set of functions, $ (X_{1},m(X_{1})),\ldots,(X_{L_{n}},m(X_{L_{n}})) $ is a set of input values evaluated with the computer model of size $ L_{n} \in \N $ and $ pen_{n}^2(\cdot) $ is a penalty term which usually penalizes the roughness of the function and which is positive for each $ f \in \F $, i.e. $ pen_{n}^2(f) \geq 0 $.
If the input dimension is smaller or equal to $ 3 $ than for example smoothing spline estimates can be applied as shown in \cite{KoKr2017a}. For bigger input dimensions neural network estimates can be used as in \cite{GoKeKo2018} and \cite{KeKo2019}.

As discussed, usually every computer model has an inherent error. 
To improve the surrogate model an estimator of the residuals can be constructed by first calculating the residuals of the surrogate model on the experimental data
\begin{equation}
\label{eq:residuals}
\epsilon_{i} = Y_{i} - \hat{m}_{L_{n}}(X_{i}) 
\quad
(i = 1,\ldots,n)
\end{equation}
and by applying then a (penalized) least squares estimate to this sample, i.e. by computing
\begin{equation}
\label{eq:freq_surrogate_residuals}
\hat{m}_{n}^{\epsilon}(\cdot)
= 
\arg \min_{f \in \bar{\F}  } 
\frac{1}{n} \sum_{i = 1}^{n} |f(X_{i}) - \epsilon_{i} |^2
+ pen_{n}^2(f),
\end{equation}
where $ \bar{\F} $ is a set of functions and $ pen_{n}^2(\cdot) $ is a penalty term for which $ pen_{n}^2(f) \geq 0 $ holds, for every $ f \in \bar{\F} $.

In the last step, the improved surrogate model is defined by 
\begin{equation}
\label{eq:freq_improved_surrogate}
\hat{m}_{n}(x) = \hat{m}_{L_{n}}(x) + \hat{m}_{n}^{\epsilon}(x) 
\quad (x \in \R). 
\end{equation}

In case that only a small sample of experimental data is available, i.e. $ n $ is small, the residual estimator usually does not yield sufficient results. Then one can use a weighted (penalized) least squares estimate instead, defined by 
\begin{equation}
\label{eq:freq_surrogate_residuals_weighted}
\hat{m}_{n}^{\epsilon}(\cdot)
= 
\arg \min_{f \in \bar{\F}  } 
\left(
\frac{w^{(n)}}{n} \sum_{i = 1}^{n} |f(X_{i}) - \epsilon_{i} |^2
+ \frac{(1-w^{(n)})}{N_{1,n}} \sum_{i = 1}^{N_{1,n}} |f(X_{i}) - 0 |^2
+ pen_{n}^2(f)
\right),
\end{equation}
where $ w^{(n)} \in [0,1] $ is a weighting term, which should be chosen data dependent and $ X_{n+L_{n}+1},\ldots, X_{n+L_{n}+N_{1,n}} $ is a set of additional input values of size $ N_{1,n} \in \N $. 
Note that in \eqref{eq:freq_surrogate} and \eqref{eq:freq_surrogate_residuals} different sets of functions are used. In an application this is usually due to different chosen function class parameters.

\subsection{Density estimation based on an improved surrogate model}
\label{se:density_improved_surrogate}

One approach to quantifying the uncertainty in the outcome of an experiment with a technical system is to estimate the underlying density of $ Y $.
If the density $ g \colon \R \to \R $ is known, we can calculate the probability $ \PROB \{ Y \in B \} $ for any given set $ B \subseteq \R $ by
\begin{equation}
\int_{B} g(x) \;
dx.
\end{equation}
In the following section we will describe a method to construct an estimator $ \hat{g} $ of the density $ g $ of $ Y $.
A simple approach is to apply the kernel density estimate of \cite{Ro1956} and \cite{Pa1962} defined by 
\begin{equation}
g(y) = \frac{1}{n \cdot h_{n}} \cdot \sum_{i=1}^{n} K \left( \frac{y-Y_{i}}{h_{n}}  \right).
\end{equation}
Usually in an application the sample size of experimental data will be too small to achieve satisfying results.
Instead one can use an improved surrogate model as described in Subsection \ref{se:surrogate_model}.

\cite{KoKr2017a}, \cite{GoKeKo2018} and \cite{KeKo2019} proposed similar methods to construct a density estimate based on an improved surrogate model, where the main difference is that the first two assumed that the full information model is available, where the last assumed that the experimental setup model is given and that \cite{KoKr2017a} used smoothing spline estimates, where \cite{GoKeKo2018} and \cite{KeKo2019} used least squares neural network estimates to construct the improved surrogate model. 
Summarizing, the method works as follows:

Estimate an improved surrogate model as discussed in Subsection \ref{se:surrogate_model}.
Next apply a kernel density estimator to an independent and identically distributed sample of $ (X,\hat{m}_{n}(X)) $ of size $ N_{2,n} \in \N $, i.e. to
\begin{equation}
\label{eq:surrogate_density_estimator}
\hat{g}_{N_{2,n}}(y) = \frac{1}{N_{2,n} \cdot h_{N_{2,n}}} \cdot \sum_{i = 1}^{N_{2,n}} 
K \left( \frac{y- \hat{m}_{n}(X_{n+L_{n}+N_{1,n}+i})}{h_{N_{2,n}}}  \right),
\end{equation}
where $ h_{N_{2,n}} > 0 $ so-called bandwidth and $ K \colon \R \to \R $ the kernel function are parameters of the estimate.
To choose the parameter $ h_{N_{2,n}} $, a data dependent approach can be used as shown in \cite{FeKoKr2015a} or \cite{FeKoKr2015b}.

\subsection{Quantile estimation based on an improved surrogate model}
\label{se:quantile_improved_surrogate}
Another approach to quantify the uncertainty in the outcome of an experiment with a technical system is to estimate its $ \alpha $-quantile, defined by
\begin{equation}
q_{Y,\alpha} = \inf \left\{
y  \in \R : G(y) \geq \alpha
\right\},
\end{equation}
where $ \alpha \in (0,1) $ and $ G $ is the cumulative distribution function of $ Y $ given by
\begin{equation}
G(y) = \PROB \left\{ Y \leq y \right\}.
\end{equation}

A simple and straight-forward approach is to use a Monte Carlo estimator. \cite{Enetal2016} and \cite{KoKr2018a} proposed to use an estimator based on a surrogate model, where they assumed that the computer simulation fits the reality perfectly. As discussed before, a computer simulation is almost always imperfect and thus a simple modification is to use an improved surrogate model.
The estimator is then constructed as follows:

Estimate an improved surrogate model as discussed in Subsection \ref{se:surrogate_model}.
Next apply a Monte Carlo quantile estimator to an independent and identically distributed sample of $ (X,\hat{m}_{n}(X)) $ of size $ N_{2,n} \in \N $, defined by 
\begin{equation}
\hat{q}_{\hat{m}_{n}(X),N_{2,n},\alpha} = \inf \left\{ y \in \R : \hat{G}_{\hat{m}_{n}(X),N_{2,n}}(y) \geq \alpha \right\},
\end{equation}
where
\begin{equation}
\hat{G}_{\hat{m}_{n}(X),N_{2,n}}(y) = \frac{1}{N_{2,n}} \sum_{i = 1}^{N_{2,n}} I_{(-\infty,m_{n}(X_{n+L_{n}+N_{1,n}+i})]}(y).
\end{equation}
Note that this is equivalent to choosing the $ \lceil N_{2,n} \cdot \alpha \rceil $ biggest value of $ \hat{m}_{n}(X_{n+L_{n}+N_{1,n}+i}) $, $ \ldots $,$ \hat{m}_{n}(X_{n+L_{n}+N_{1,n}+N_{2,n}}) $.

\subsection{Application}
In the following, we will illustrate and discuss the above described methods by applying them to data of the technical systems described in Section \ref{se:technical_systems}. In case of the MAFDS we restrict ourselves to use only $ 10 $ randomly chosen experimental data points of the available $ 100 $.

\subsubsection{Estimation of an improved surrogate model}

Many of the in the following described methods of uncertainty quantification need an (improved) surrogate model of the computer model. If not stated otherwise the (improved) surrogate model will be estimated as follows:
In case of the MAFDS, cf. Subsection \ref{se:MAFDS}, the input dimension $ d $ is $ 1 $. Here we use in \eqref{eq:freq_surrogate} a smoothing spline estimator as implemented by the {\it MATLAB} method {\it csaps}, where the smoothing parameter is chosen by generalized cross-validation. 
In \eqref{eq:freq_improved_surrogate} the smoothing parameter and the weighting parameter $ w^{(n)} \in \{0,0.1,0.2,\ldots,1\} $ are chosen simultaneously by a $ 5 $-fold cross validation, where the empirical $ L_{2} $ risk is calculated only on the experimental data.
In case of the piezo-beam, cf. Subsection \ref{se:piezo_beam}, we use fully connected feed forward neural networks from the {\it MATLAB} {\it Deep Learning Toolbox} to estimate the surrogate model. The network topology is chosen data dependent by a splitting of the sample, where we use $ 2/3 $ training and $ 1/3 $ testing data. We use networks with $ \{1,5,10\} $ hidden layer which have $ \{1,6,11,16,21\} $ neurons per layer. 
To estimate an improved surrogate model we use a special class of sparsely connected neural networks. A detailed description of these neural networks and their implementation can be found in \cite{KeKo2019}.

\begin{figure}[h!]
	\centering
	\includegraphics[width=7cm]{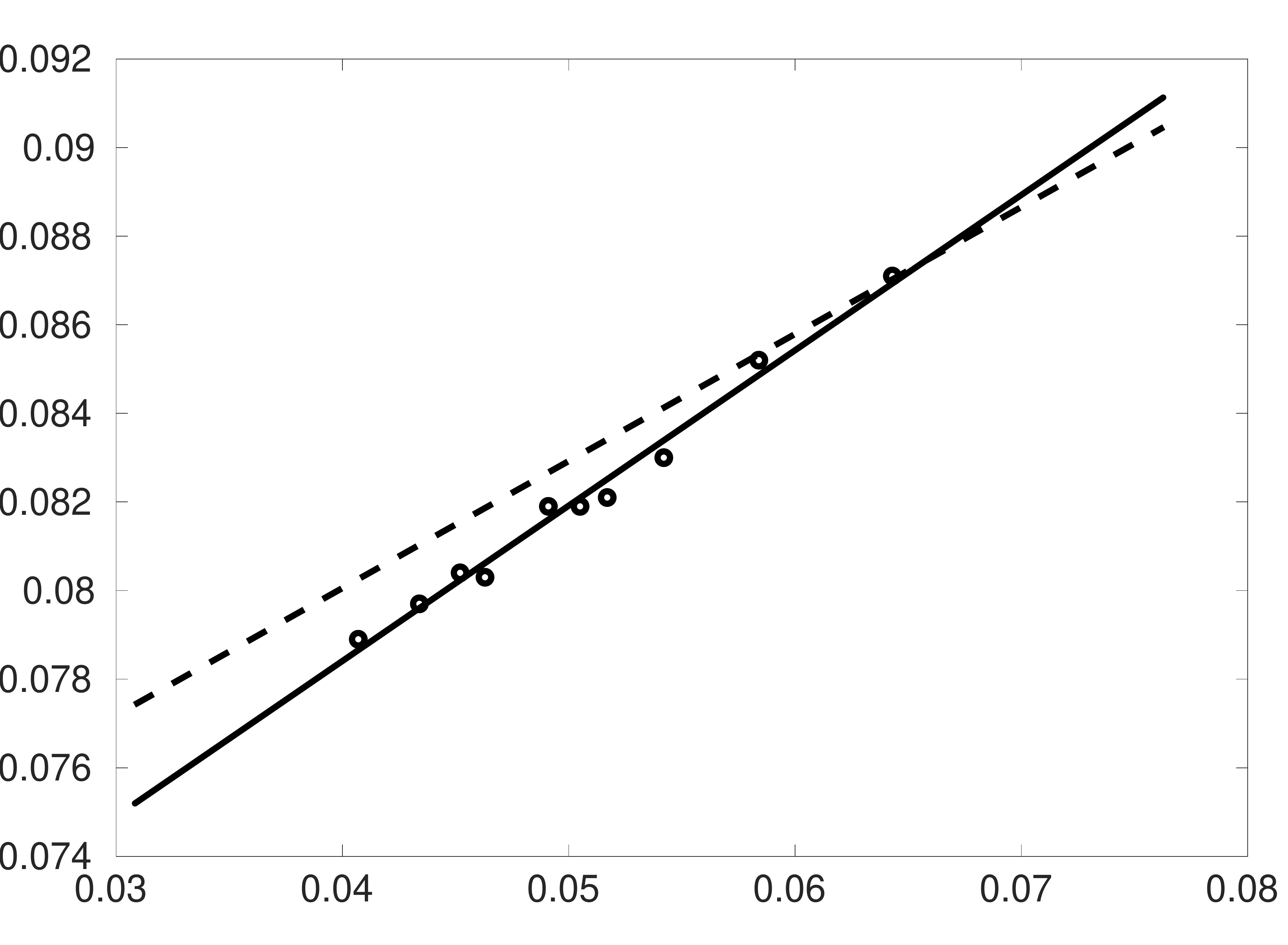}
	\includegraphics[width=7cm]{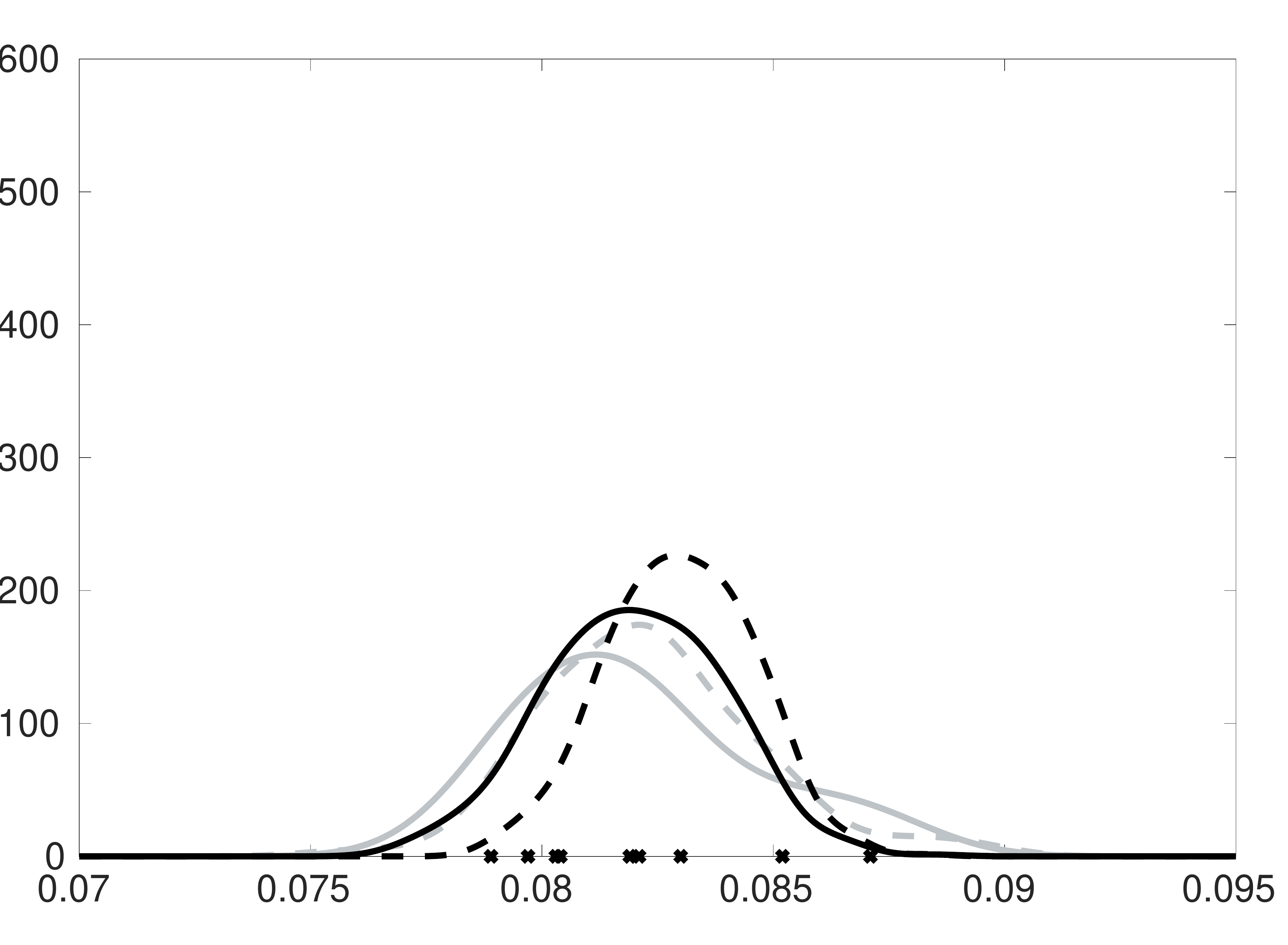}
	\caption{\label{fig:plot_MAFDS_surrogate_cm1_KoKr2017}Left plot contains $ 10 $ randomly chosen experimental data points of the MAFDS (circles) used to construct the improved surrogate model, the surrogate model based on computer model 1 (dashed line) and the improved surrogate model (solid line). Right plot contains a kernel density estimator based on the $ 10 $ experimental data (gray solid line), a kernel density estimator based on the computer model 1 (black dashed line) and kernel density estimator based on an improved surrogate model (black solid line). The $ 10 $ experimental data points are indicated on the x-axis. Furthermore, for comparison a kernel density estimator based on all available experimental data (gray dashed line).}
\end{figure}

\begin{figure}[h!]
	\centering
	\includegraphics[width=7cm]{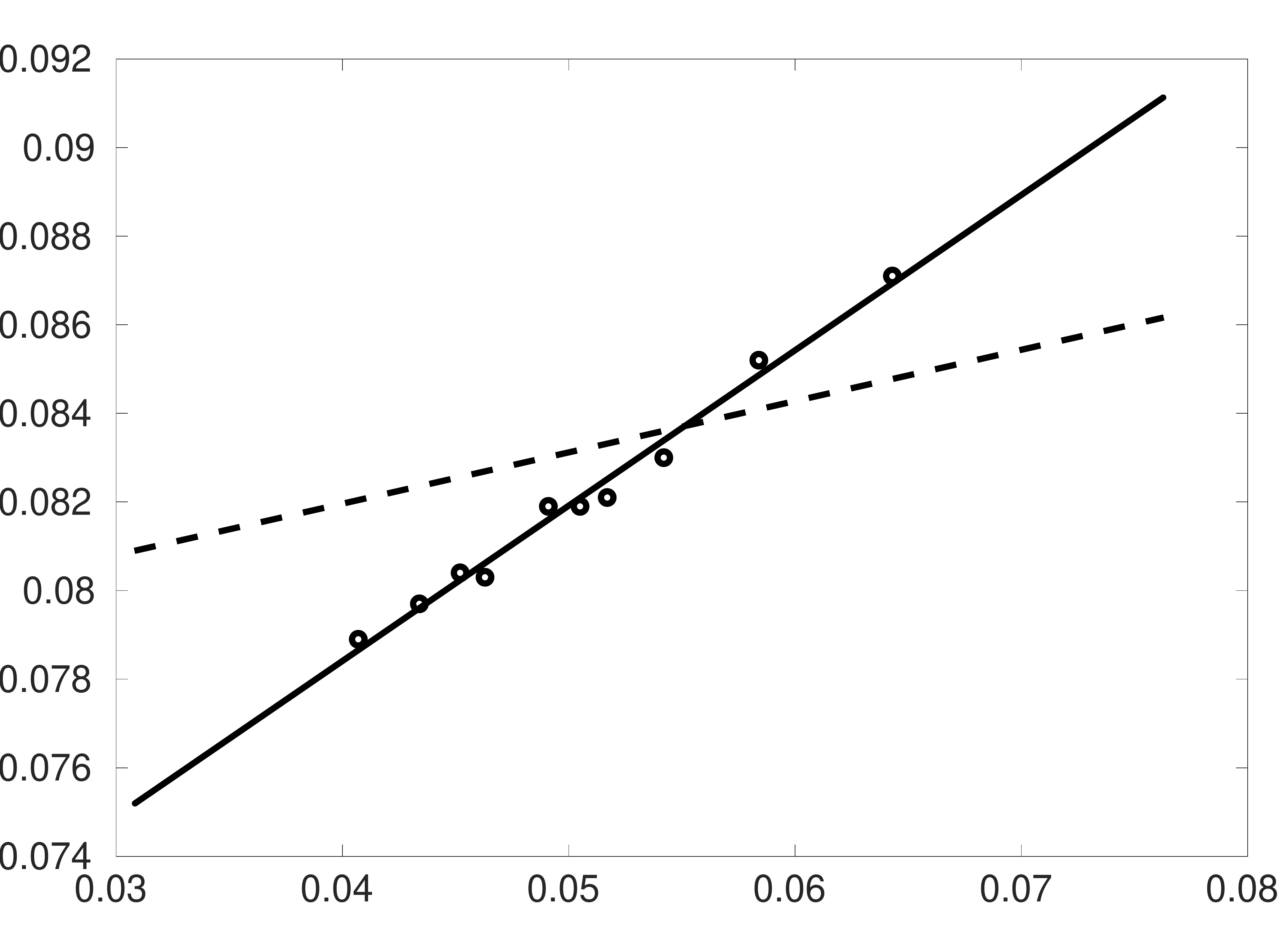}
	\includegraphics[width=7cm]{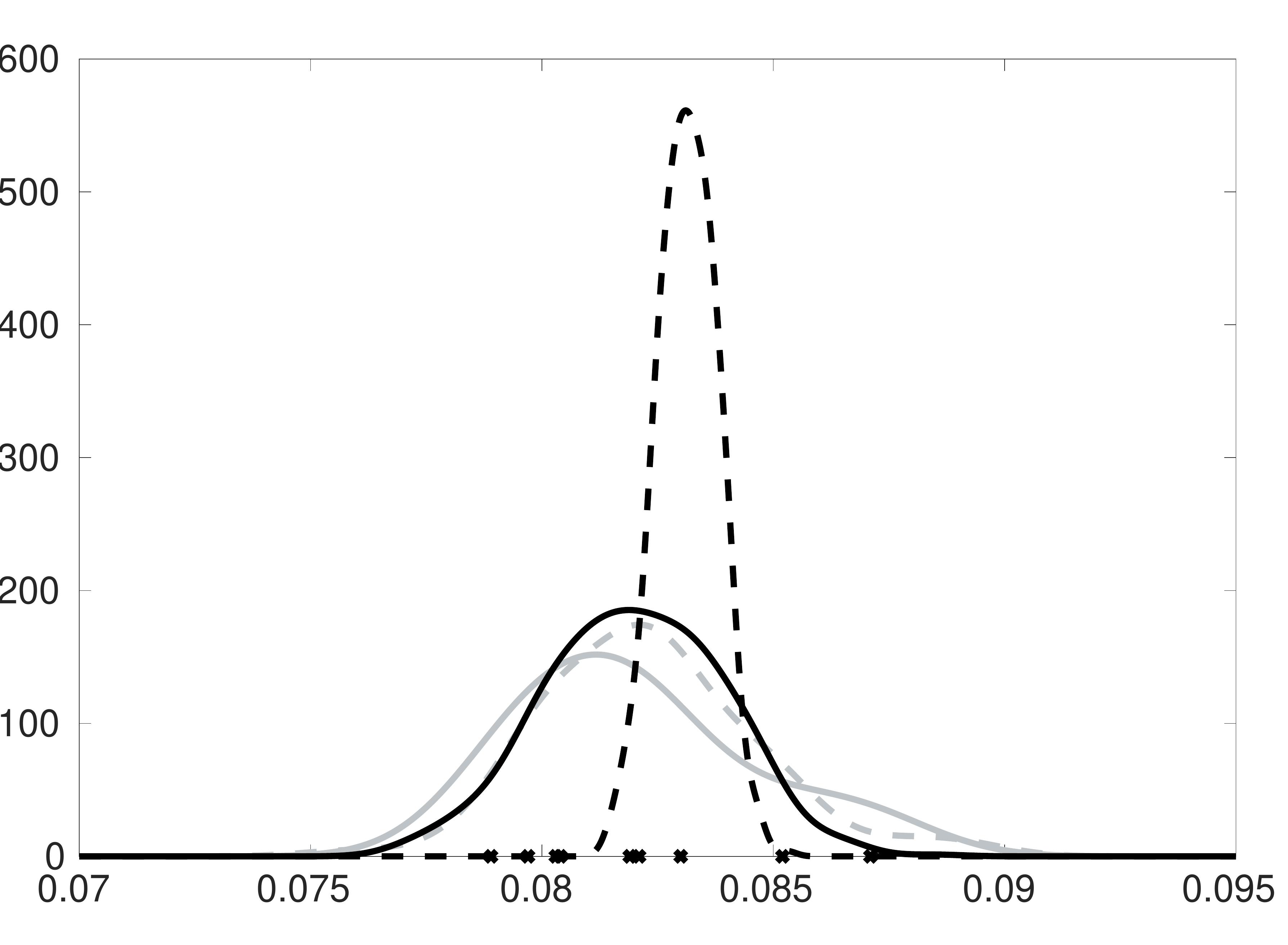}
	\caption{\label{fig:plot_MAFDS_surrogate_cm2_KoKr2017}Left plot contains $ 10 $ randomly chosen experimental data points of the MAFDS (circles) used to construct the improved surrogate model, the surrogate model based on computer model 2 (dashed line) and the improved surrogate model (solid line). Right plot contains a kernel density estimator based on the $ 10 $ experimental data (gray solid line), a kernel density estimator based on the computer model 2 (black dashed line) and kernel density estimator based on an improved surrogate model (black solid line). The $ 10 $ experimental data points are indicated on the x-axis. Furthermore, for comparison a kernel density estimator based on all available experimental data (gray dashed line).}
\end{figure}

For the MAFDS the resulting improved surrogate models, together with the experimental data and the surrogate models based on the two computer models are displayed in the left plots of Figures \ref{fig:plot_MAFDS_surrogate_cm1_KoKr2017} and \ref{fig:plot_MAFDS_surrogate_cm2_KoKr2017}.
The method estimates (nearly) identical surrogate models for both computer simulation. The resulting surrogate model fits the experimental data better than both computer simulations.

Since the piezo-beam depends on $ 5 $ input parameters, we are not able to visualize the result for the piezo-beam.

\subsubsection{Density estimation based on an improved surrogate model}

\begin{figure}[h!]
	\centering
	\includegraphics[width=7cm]{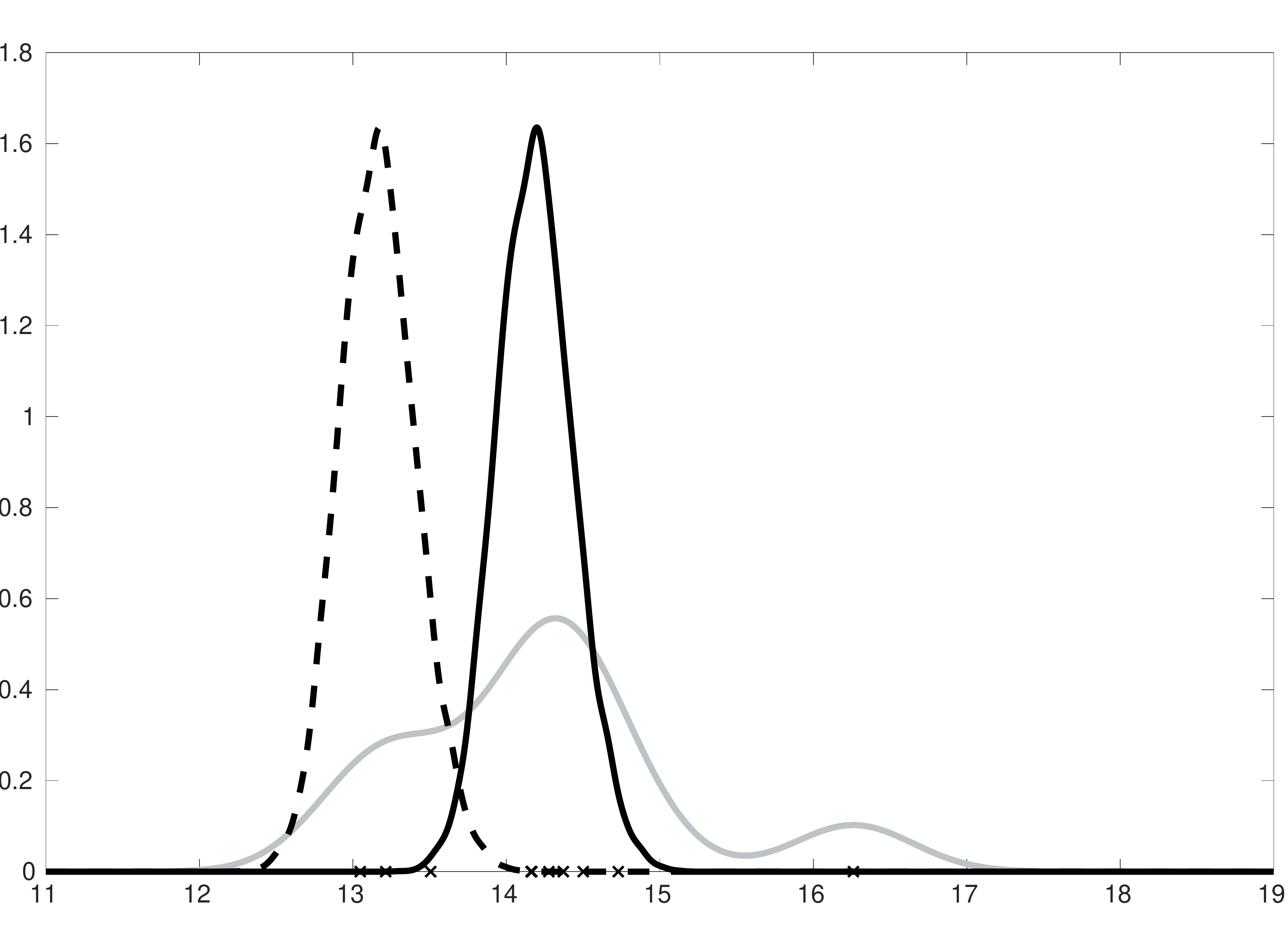}
	\caption{\label{fig:plot_KoKr2017}Plot compares density estimator of piezo-beam, cf. Subsection \ref{se:piezo_beam}. A kernel density estimator based on all available experimental data (gray solid line), a kernel density estimator based on the computer model (black dashed line) and kernel density estimator based on an improved surrogate model (black solid line). The values of the outcome of the $ 10 $ experiments are indicated on the x-axis.}
\end{figure}

We apply the density estimation method to data of both technical systems.
The surrogate models are constructed as described in Subsection \ref{se:surrogate_model}.
The results are displayed in Figures \ref{fig:plot_MAFDS_surrogate_cm1_KoKr2017}, \ref{fig:plot_MAFDS_surrogate_cm2_KoKr2017} and \ref{fig:plot_KoKr2017}. 
In case of the MAFDS, the resulting improved surrogate models are (almost) equal for both computer models, although the estimators based solely on the computer models differ a lot. For both computer models, the improved surrogate model seems to correct their prediction error and yields a more accurate estimation of the real density. 
In case of the piezo-beam, the resulting improved surrogate model predict higher values than the computer simulation, which fits the experimental data better. 

To conclude, in both applications the improved surrogate model is able to correct the computer model error and yields a better prediction of the real experiments, resulting in an density estimation which fits the reality better than the density estimation based on computer simulations, or the density estimator based solely on experimental data.

\subsubsection{Quantile estimation based on an improved surrogate model}
We apply the quantile estimation method described in Subsection \ref{se:quantile_improved_surrogate} to the data of both technical systems.
The surrogate models are constructed as described in Subsection \ref{se:surrogate_model}.
In case of the MAFDS we restrict ourselves to use only $ 10 $ of the available $ 100 $ experimental data points to estimate the improved surrogate model. 
For both technical systems we are interested in the $ 0.95 $-quantile. The resulting estimates are shown in Table  \ref{ta:results_quantiles}.
\begin{table}[h] 
\hspace*{-2cm}\centering 
\begin{tabular}{c|c|c}
	& MAFDS & piezo-beam  \\ 
	computer model 1 & $ 0.0849 $ & $ 14.6994 $  \\
	computer model 2 & $ 0.0849 $ &   \\ 
\end{tabular} 
\caption{\label{ta:results_quantiles} Results of the Monte Carlo estimator based on an improved surrogate model of the $ 0.95 $-quantile of the outcome $ Y $ of the experiments with both technical systems.} 
\end{table} 
As discussed in the last section in case of the MAFDS, the resulting improved surrogate models are (almost) equal for both computer models and thus the resulting quantile estimate is (almost equal).

\section{Comparison of computer models}
\label{se:computer_model_comparision}
In this section methods which can be used to compare different computer simulations are described.
\subsection{Comparison of computer models via cumulative distribution functions}
\label{se:AVM}

\cite{RoOb2011} introduced a method to detect model form uncertainty. The term model form is in their context the result of all assumptions, conceptualizations, abstractions, approximations, and mathematical formulations on which the model relies. Consequently an imperfect model is from their perspective the result of wrong assumptions in the model form. To detect uncertainty in the model form, they proposed the so-called Area Validation Metric (AVM). The metric compares the cumulative density function of numerical simulated data and of experimental measurements. It uses a sample of experimental and computer simulated data
\begin{equation}
\label{se4eq1}
Y_1,\ldots ,Y_n
\quad \text{and} \quad
m(X_{n+1}),\ldots,m(X_{n+L_{n}})
\end{equation}
of size $ n \in \N $ and $ L_{n} \in \N $, where as described in Section \ref{se:data_models}, $ m \colon \Rd \to $ is a computer simulation which is designed to emulate an experiment with a technical system and $ X \in \Rd $ is the corresponding input for the experiment and $ Y \in \R $ is the output.
Based on \eqref{se4eq1} the cumulative density functions are estimated by the empirical distribution function defined by
\begin{equation}
\label{}
\hat{F}_{Y}(t) = \frac{1}{n} \sum_{i=1}^{n} I_{(-\infty,Y_{i}]}(t)
\quad \text{and} \quad
\hat{F}_{m(X)}(t) = \frac{1}{L_{n}} \sum_{i=n+1}^{n+L_{n}} I_{(-\infty,m(X_{i})]}(t).
\end{equation}
Finally the AVM is calculated by
\begin{equation} 
\label{eq:avm}
\int | \hat{F}_{Y}(t) - \hat{F}_{m(X)}(t) | dt.
\end{equation}

\subsection{A Bayesian approach for quantifying the computer model error}
\label{se:KOH}

\cite{KeOh2001} proposed a Bayesian method for the calibration of computer models in case that the
underlying mathematical model is imperfect. They assume that their data is given by
\begin{equation}
\label{se2eq1}
Y=m(\theta,x)+\delta(x)+\epsilon
\end{equation}
where $m(\cdot,\cdot)$ describes the computer model depending on some parameter $\theta$ and some vector $x$ describing uncertain factors of the reality, where $\delta$ is the model error and $\epsilon$ is a normally distributed error with expectation zero which contains for instance measurement errors. They propose Bayesian techniques for the choice of the optimal parameter value $\theta^*$ (so called calibration), which also model the model error $\delta$ and are therefore able to quantify the model error of the
computer model. In the sequel we assume that our system is already calibrated in the sense that our computer model is based on some physical model for which we have adjusted the parameters already to reality based on separate experiments with our technical system. In this case the model of \cite{KeOh2001} simplifies to
\begin{equation}
\label{se2eq2}
Y = m(x)+\delta(x)+\epsilon.
\end{equation}
Here $m(\cdot)$ is the (already calibrated) computer model, $\delta$ is the discrepancy term of this model, and $\epsilon$ is a random normally distributed error with expectation zero and variance $ \lambda $.
In addition  $x$ is (a possible random) value from $\Rd$ describing input parameters of the experiment with the technical 
system (e.g., drop height in experiments with the MAFDS, cf. Subsection \ref{se:MAFDS}).

From model \eqref{se2eq2} it is assumed that $n$ observations are observed, i.e., the observed data is 
\begin{equation}
\label{se2eq3}
(X_1,m(X_1),Y_1), \dots, (X_n,m(X_n),Y_n)
\end{equation}
where
\begin{equation}
\label{se2eq4}
Y_i = m(X_i)+\delta(X_i)+\epsilon_i \quad (i=1, \dots, n).
\end{equation}
Here $\epsilon, \epsilon_1, \dots, \epsilon_n$
are independent and identically distributed.
\cite{KeOh2001} uses a Gaussian process with mean function $\mu(\cdot):\Rd \rightarrow \R$ and covariance function $c(\cdot,\cdot):\Rd \times \Rd \rightarrow \R_+$ to model the discrepancy term $\delta(\cdot)$.
This means that for any $k \in \N$ and any $X_1, \dots, X_k \in \Rd$ 
the vector
\[
(\delta(X_1),\delta(X_2), \dots, \delta(X_k))^T
\]
is multivariate normally distributed with mean vector $(\mu(X_1), \dots, \mu(X_k))^T$
and covariance matrix $(c(X_i,X_j))_{1 \leq i,j \leq k}$.
The random error is chosen normally distributed with mean zero
and variance $\lambda$. So as soon as
$\lambda$, the mean function $\mu(\cdot)$ and the covariance function $c(\cdot,\cdot)$
are chosen, the model \eqref{se2eq2} is completely specified
for given $x$.

In order to specify $\mu(\cdot)$ and $c(\cdot,\cdot)$, \cite{KeOh2001} propose to use a hierarchical Bayesian model. Here the choice of $\mu(\cdot)$ and $c(\cdot,\cdot)$ could be based on prior knowledge about the technical system, but since this is difficult to incorporate in this model they suggest to use the following simple choice: 
First prior densities for the parameters of the hierarchical Bayesian model are chosen from a parametric model. Therefore prior knowledge (of the technical system) can be used to choose these densities, or if such knowledge is not available, normal distributions or a choice of the form 
\begin{equation}
\label{se2eq5}
p(t) = c \cdot \frac{1}{t},
\end{equation} 
with parameter $ c \in \R_{+} $ and where $p(t)$ must be set to zero for large $t$, is recommended. 

In a second step they choose $\mu(\cdot)$ as a constant function
\begin{equation}
\label{se2eq6}
\mu(x) = \beta \quad (x \in \Rd)
\end{equation}
and use for $c(\cdot,\cdot):\Rd \times \Rd \rightarrow \R$ the definition $ c(z_1,z_2) = \sigma^2 R(z_{1}.z_{2}) $, where 
\begin{equation}
\label{se2eq7}
R(z_1,z_2) = \exp \left(
- \sum_{j=1}^d \omega_j \cdot (z_1^{(j)}-z_2^{(j)})^2
\right)
\quad (z_1,z_2 \in \Rd).
\end{equation}
As prior distributions they recommend normal distributions for $\lambda$ and $\beta$, with mean and variance parameters $ \mu_{\lambda},\sigma^2_{\lambda},\mu_{\beta}$ and $\sigma^2_{\beta} $, and for all other parameters they recommend to use 
\eqref{se2eq5}, with parameters $ c_{\sigma^2},c_{\omega_{1}},\ldots,c_{\omega_{d}} $.
Thus the hierarchical Bayesian model uses parameters
\begin{equation}
\P^{(1)} = \{ \lambda,\beta,\sigma,\omega_{1},\ldots,\omega_{d} \},
\end{equation}
where the vector of parameters is a realization of a random variable
\begin{equation}
(\Lambda,B,\Sigma,\Omega_{1},\ldots,\Omega_{d}),
\end{equation}
with parametric densities 
\begin{equation}
p_{\Lambda}(\cdot),\ldots,p_{\Omega_{d}}(\cdot),
\end{equation}
with distribution parameters 
\begin{equation}
\P^{(2)} = \{ \mu_{\Lambda}, \sigma^2_{\Lambda}, \mu_{B}, \sigma^2_{B}, c_{\Sigma^2}, c_{\Omega_{1}}, \ldots, c_{\Omega_{d}} \}.
\end{equation}
\cite{KeOh2001} recommended to choose fixed values for the parameters in $ \P^{(2)} $ whenever possible, e.g. from information about the physical system, to reduce complexity.

The next step is to compute the posterior density of $ (\Lambda,B,\Sigma,\Omega_{1},\ldots,\Omega_{d}) $ given the data \eqref{se2eq3}. From the
above definitions we get that the conditional density
of $(Y_1, \dots, Y_n)$ given $ (\Lambda,B,\Sigma,\Omega_{1},\ldots,\Omega_{d}) = (\lambda,\beta,\sigma,\omega_{1},\ldots,\omega_{d}) $  is given by
\begin{eqnarray}
\label{se2eq9}
&&
\hspace{-5mm}
f_{(Y_1,\dots,Y_n)|(\Lambda,B,\Sigma,\Omega_{1},\ldots,\Omega_{d}) = (\lambda,\beta,\sigma,\omega_{1},\ldots,\omega_{d}) }(y_1, \dots, y_n)
=  
\nonumber
\\
&& 
\quad \frac{1}{\sqrt{
(2 \pi)^n \cdot det(\Theta)
}}
\cdot
\nonumber
\exp
\Big(
- \frac{1}{2}
\cdot
(y_1-m(X_{1})-\beta, \dots, y_n -m(X_{n}) - \beta)
\cdot
\Theta^{-1}
\cdot
\nonumber
\\
&& \hspace{55mm} (y_1-m(X_{1})-\beta, \dots, y_n -m(X_{n}) - \beta)^T
\Big)
\end{eqnarray}
where
\[
\Theta
=
\left(
c(X_k,X_l) + \lambda \cdot 1_{\{k=l\}}
\right)_{1 \leq k,l \leq n},
\]
from which we can conclude that we have
\begin{eqnarray}
\label{se2eq10}
&&
\hspace{-6mm} 
 f_{(Y_1, \dots, Y_n)}(y_1, \dots, y_n) =
\\
\nonumber
&& \hspace*{-5mm} \int \!\! \ldots \!\! \int
f_{(Y_1,\dots,Y_n)|(\Lambda,B,\Sigma,\Omega_{1},\ldots,\Omega_{d})=(\lambda,\beta,\sigma,\omega_{1},\ldots,\omega_{d})}(y_1, \dots, y_n)
\cdot p_{\Lambda}(\lambda) \cdots p_{\Omega_d}(\omega_d)
\, d \lambda \dots d \omega_d
\end{eqnarray}
and
\begin{eqnarray}
\label{se2eq11}
&& 
\hspace{-5mm} 
f_{(\Lambda,B,\Sigma,\Omega_{1},\ldots,\Omega_{d})|(Y_1,\dots,Y_n) = (y_1,\dots,y_n)}(\lambda, \dots, \omega_d)
=
\nonumber
\\
&& \hspace{5mm} 
\frac{
	f_{(Y_1,\dots,Y_n)|(\Lambda,B,\Sigma,\Omega_{1},\ldots,\Omega_{d}) = (\lambda,\beta,\sigma,\omega_{1},\ldots,\omega_{d}) }(y_1, \dots, y_n)
	\cdot p_{\Lambda}(\lambda) \cdot \dots \cdot p_{\Omega_d}(\omega_d)
}{
f_{(Y_1, \dots, Y_n)}(y_1, \dots, y_n)
},
\end{eqnarray}
where $ p_{\Lambda}(\cdot),\ldots,p_{\Omega_{d}}(\cdot) $, are the above specified prior densities with the corresponding parameters $ \P^{(2)} $ .
Note that the value of \eqref{se2eq11} also depends on the values of the distribution parameters $ \P^{(1)} $.

\cite{KeOh2001} propose next to estimate the parameters $  \P^{(1)} \cup \P^{(2)} $ by a maximum likelihood estimate and threat the estimated values as if they were the real values, i.e. they proposed to compute 
\begin{eqnarray}
\label{eq:maximum_likelihood}
\hat{\P}^{(1)} \cup \hat{\P}^{(2)}
=
\arg \max_{ \P^{(1)} \cup \P^{(2)} }
f_{(\Lambda,B,\Sigma,\Omega_{1},\ldots,\Omega_{d})|(Y_1,\dots,Y_n)= (y_1,\dots,y_n)}(\lambda, \dots, \omega_d),
\end{eqnarray}
where $ \hat{\P}^{(1)} \cup \hat{\P}^{(2)} $ are the resulting estimates of all parameters included in $ \P^{(1)} \cup \P^{(2)} $.
To compute this estimate one can use a numerical (approximative) maximation of the posterior density (where a maximum with respect
to $2d+8$ variables has to be computed). 
To reduce the complexity in solving \eqref{eq:maximum_likelihood} one can also complete the square for $ B $ in \eqref{se2eq9}
to find 
\begin{equation*}
\hat{\beta} = \frac{(1,\ldots,1) \cdot \Theta^{-1} \cdot (y_{1}-m(X_{1}),\ldots,y_{n}-m(X_{n}))^T }{(1,\ldots,1) \cdot \Theta^{-1} \cdot (1,\ldots,1)^T}.
\end{equation*}
Then the above maximum has only be computed with respect to $ 2d+5 $ variables, since if $ \hat{\beta} $ is known, we do not need to estimate it's distribution parameters $ \mu_{B},\sigma_{B}^2 $.
Alternatively one can simplify the computation by choosing for some of the above variables plausible estimates, e.g., $ B $ can be estimated by
\begin{equation}
\label{eq:empirical_beta}
\hat{\beta} = \frac{1}{n} \sum_{i=1}^n (Y_i - m(X_i)).
\end{equation}

As soon as we have these estimate we know the distribution
of the error
\begin{equation*}
(Y_i - m(X_i))_{i=1, \dots,n}= (\delta(X_i)+ \epsilon_i)_{i=1,\dots,n},
\end{equation*}
which is a multivariate normal distribution with mean vector
$(\hat{\beta}, \dots, \hat{\beta})^T$ and covariance matrix
\begin{equation*}
\left(
\hat{\sigma}^2 \cdot \exp \left(
- \sum_{j=1}^d \hat{\omega}_j \cdot (X_k^{(j)}-X_l^{(j)})^2
\right)
+ \hat{\lambda} \cdot 1_{\{k=l\}}
\right)_{1 \leq k,l \leq n}.
\end{equation*}
From this result we can, e.g., estimate quantiles of the absolute
model error
\begin{equation*}
(|\delta(X_i)+ \epsilon_i|)_{i=1,\dots,n}
\end{equation*}
occurring in the data \eqref{se2eq3} via Monte Carlo in order to quantify the model error 
of our simulation model.

\subsection{Quantifying the computer model error via bootstrap}
\label{se:Wong}

\cite{WoStLe2017} proposed the following method for quantifying the the model error using a bootstrap approach: 

First estimate a surrogate model $ \hat{m}_{L_{n}} $ by \eqref{eq:freq_surrogate} as described in Subsection \ref{se:surrogate_model}. Next to quantify the model error, $ B $ bootstrap samples are generated. Therefore compute the residuals as in Subsection \ref{se:surrogate_model} by \eqref{eq:residuals} and generate the bootstrap samples $(X_{1,b},\epsilon_{1,b})$, $ \dots $, $(X_{n,b},\epsilon_{n,b})$ $ (b = 1 \ldots,B) $ by sampling randomly and {\bf with replacement} $ n $ values of $ (X_{1},\epsilon_{1}) $, $ \ldots $, $ (X_{n},\epsilon_{n}) $.

Next we split the bootstrap samples in two parts, a learning and an evaluation sample. Therefore we choose a $ n_{l} \in \{1,\ldots,n-1 \} $. Then we construct $ B $ estimators of the residuals $ \hat{m}_{n_{l},b}^{\epsilon} $ $ (b = 1 \ldots,B) $ by \eqref{eq:freq_surrogate_residuals}, where we replace $ (X_{1},\epsilon_{1}) $, $ \ldots $, $ (X_{n},\epsilon_{n}) $ by $(X_{1,b},\epsilon_{1,b})$, $ \dots $, $(X_{n_{l},b},\epsilon_{n_{l},b})$. 
Since we only use a part of the experimental data to construct the residual estimators this can lead to unsatisfying results, if the sample size is too small as discussed in Subsection \ref{se:surrogate_model}. In this case we recommend to use the weighted version of the estimator instead, as defined in \eqref{eq:freq_surrogate_residuals_weighted}.

For each bootstrap sample we estimate the $ \alpha $-quantile $ q_{|\epsilon|,\alpha} $ of the absolute model error on the remaining input values evaluated with the bootstrap residual estimator. The $ \alpha $-quantile estimator is then defined by
\begin{equation}
\label{eq:bootstrap_quantile}
\hat{q}_{|\hat{m}_{n_{l},b}^{\epsilon}(X)|,n-n_{l},\alpha} = \min \left\{ y \in \R :  \frac{1}{n-n_{l}} \sum_{i=n_{l}+1}^{n} I_{\{ |\hat{m}_{n_{l},b}^{\epsilon} (X_{i})| \leq y\}} \geq \alpha \right\}.
\end{equation} 
With this estimator one can compare two computer models by comparing the median of their bootstrap $ \alpha $-quantiles.

\subsection{Application}
In the following, we will illustrate and discuss the above described methods by applying them to data of the MAFDS described in Subsection \ref{se:technical_systems}. Although it might be possible to apply the methods to data of the piezo-beam, cf. Subsection \ref{se:piezo_beam}, we only have one computer model available for this system and thus can not compare different computer models.

\subsubsection{Comparison of computer models via cumulative distribution functions}
We apply the in Subsection \ref{se:AVM} described method to experimental and simulated output data from the MAFDS. Data generated by different computer models is applied separately.

\begin{figure}[h!]
	\centering
	\includegraphics[width=7cm]{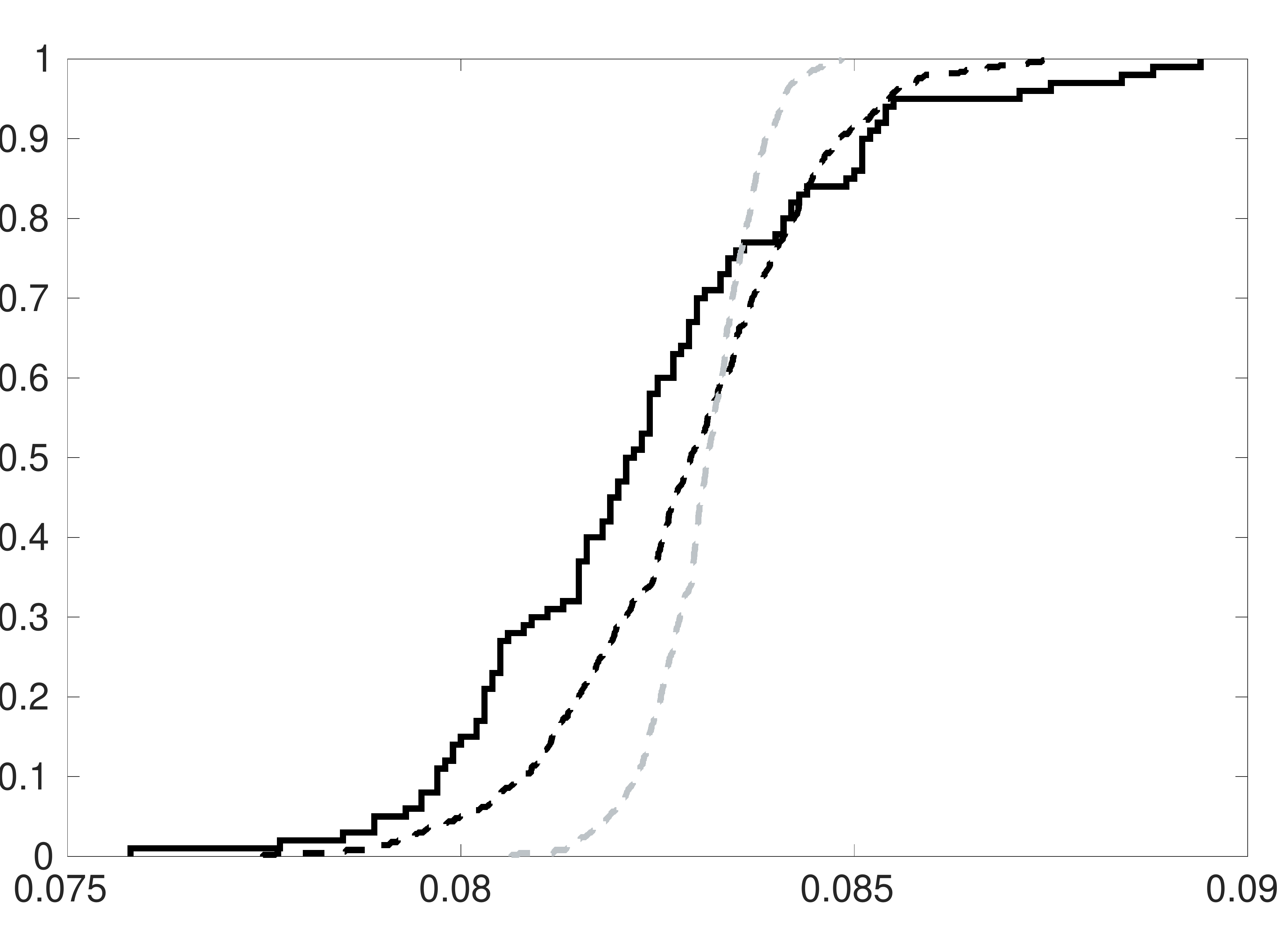}
	\caption{\label{fig:plot_RoOb2011}Three estimates of the cumulative density function of the experimental outcome of the MAFDS. An estimator based on experimental data (black solid line), an estimator based on computer experiments with computer model 1 (black dashed line) and computer model 2 (gray dashed line).}
\end{figure}

The integral in \eqref{eq:avm} is approximated via a Riemann sum calculated on equidistant points. 
The resulting values for both systems can be found in Table \ref{ta:results_RoOb2011}. 
\begin{table}[h] 
	\hspace*{-2cm}\centering 
	\begin{tabular}{c|c}
		& MAFDS   \\ 
		computer model 1 &$ 0.00084 $  \\ 
		computer model 2 &$ 0.00152 $  \\ 
	\end{tabular} 
	\caption{\label{ta:results_RoOb2011} Value of the area validation metric for both computer models of the MAFDS.} 
\end{table} 
A plot of the resulting estimates of the cumulative density functions for both technical systems is shown in Figure \ref{fig:plot_RoOb2011}.
We can use the method to compare the two different computer models of the MAFDS.
The value of the AVM is smaller for computer model 1 and also the estimated cumulative density function based on model 1 fits the estimate of the cumulative density function based on the experimental data better, cf. Figure \ref{fig:plot_RoOb2011}.
Thus based on the AVM one can conclude that model 1 predicts the reality better than model 2.

The method can be used to compare different computer simulations, but otherwise not much conclusions can be drawn, since the value of the AVM is hard to interpret on its own.

\subsubsection{A Bayesian approach for quantifying the computer model error}
We apply the in Subsection \ref{se:KOH} described method of \cite{KeOh2001} on data of the MAFDS.
The value of $ \beta $ is estimated via \eqref{eq:empirical_beta}. The remaining parameters are estimated by the maximum likelihood estimator defined by \eqref{eq:maximum_likelihood}, where $ \beta $ takes the fixed value estimated before and to approximate the minimum we use a interior-point method, as implemented by the {\it MATLAB} function {\it fmincon}. 

\begin{figure}[h!]
	\centering
	\includegraphics[width=7cm]{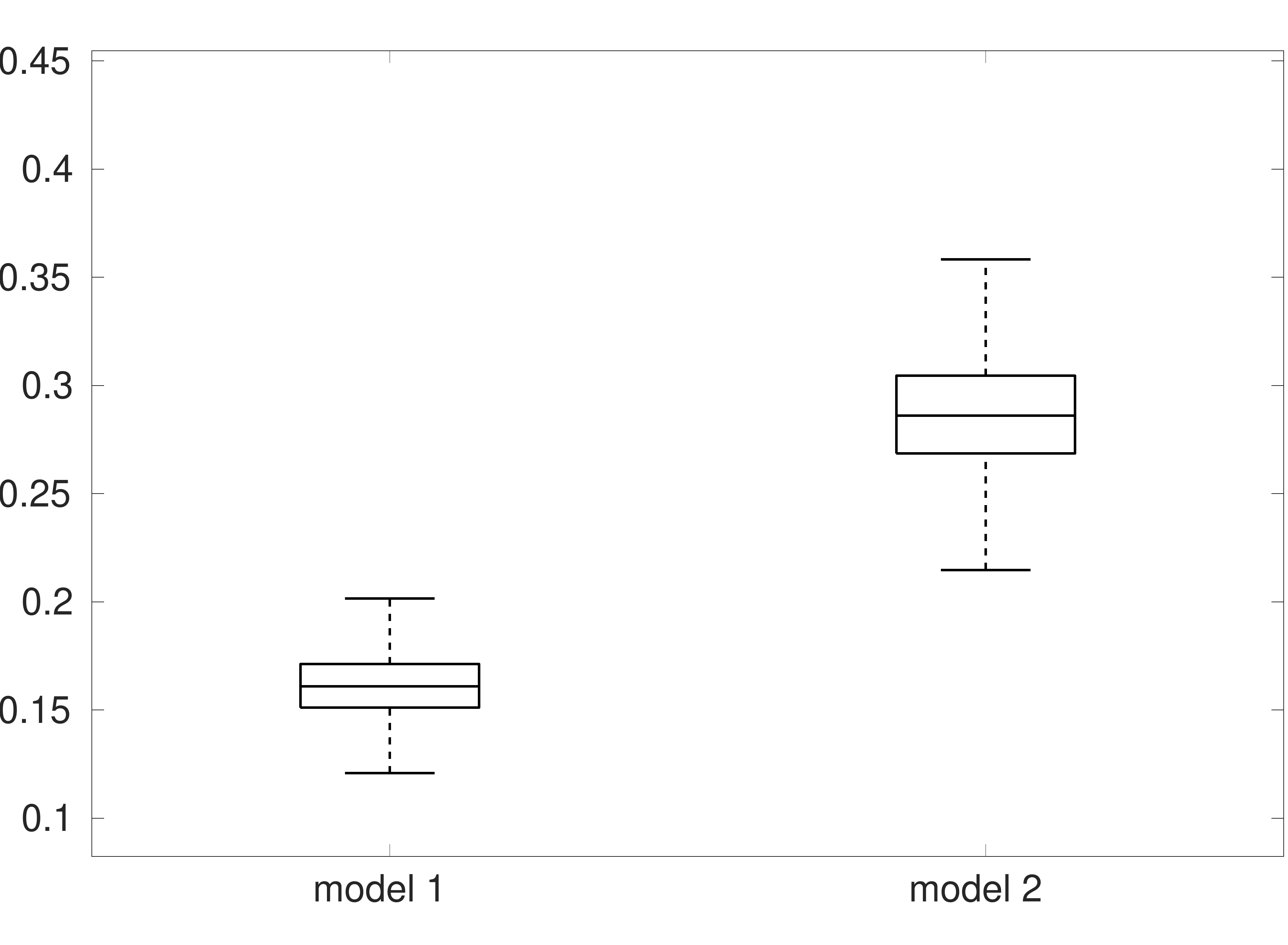}
	\caption{\label{fig:Boxplot_KeOh2001}Box plot without outliers of $ 10^6 $ estimates for the $ 0.95 $-quantiles of the $ n=100 $ absolute model errors for both computer models of the MAFDS.}
\end{figure}

We are interested in the $ 0.95 $-quantiles of the $ n = 100 $ absolute model errors. 
Therefore we generate a sample of $ 10^6 $ realizations of the $ n=100 $ model errors, using the estimated distribution parameters,  calculate the empirical $ 0.95 $-quantiles and compute their median. The results are shown in Table \ref{ta:results_KeOh2001}. 
\begin{table}[h] 
\hspace*{-2cm}\centering 
	\begin{tabular}{c|c}
		& MAFDS   \\ 
		computer model 1 & $ 0.161 $ \\
		computer model 2 & $ 0.286 $ \\ 
	\end{tabular} 
\caption{\label{ta:results_KeOh2001} Estimated $ 0.95 $-quantiles of the $ n=100 $ maximal model errors for both computer models of the MAFDS. } 
\end{table} 
Figure \ref{fig:Boxplot_KeOh2001} shows two box plots of $ 10^6 $ estimates of the $ 0.95 $-quantiles of the $ n=100 $ absolute model errors for both technical systems. 

In case of the MAFDS, we can use the method to compare the different computer models and analyse their computer model error. 
The predicted computed model error is smaller for computer model 1.
Thus based on this method one can conclude that model 1 predicts the reality better.

To conclude, the method can be used to compare different computer simulations and estimate the model error.

\subsubsection{Quantifying the computer model error via bootstrap}

\begin{figure}[h!]
	\centering
	\includegraphics[width=7cm]{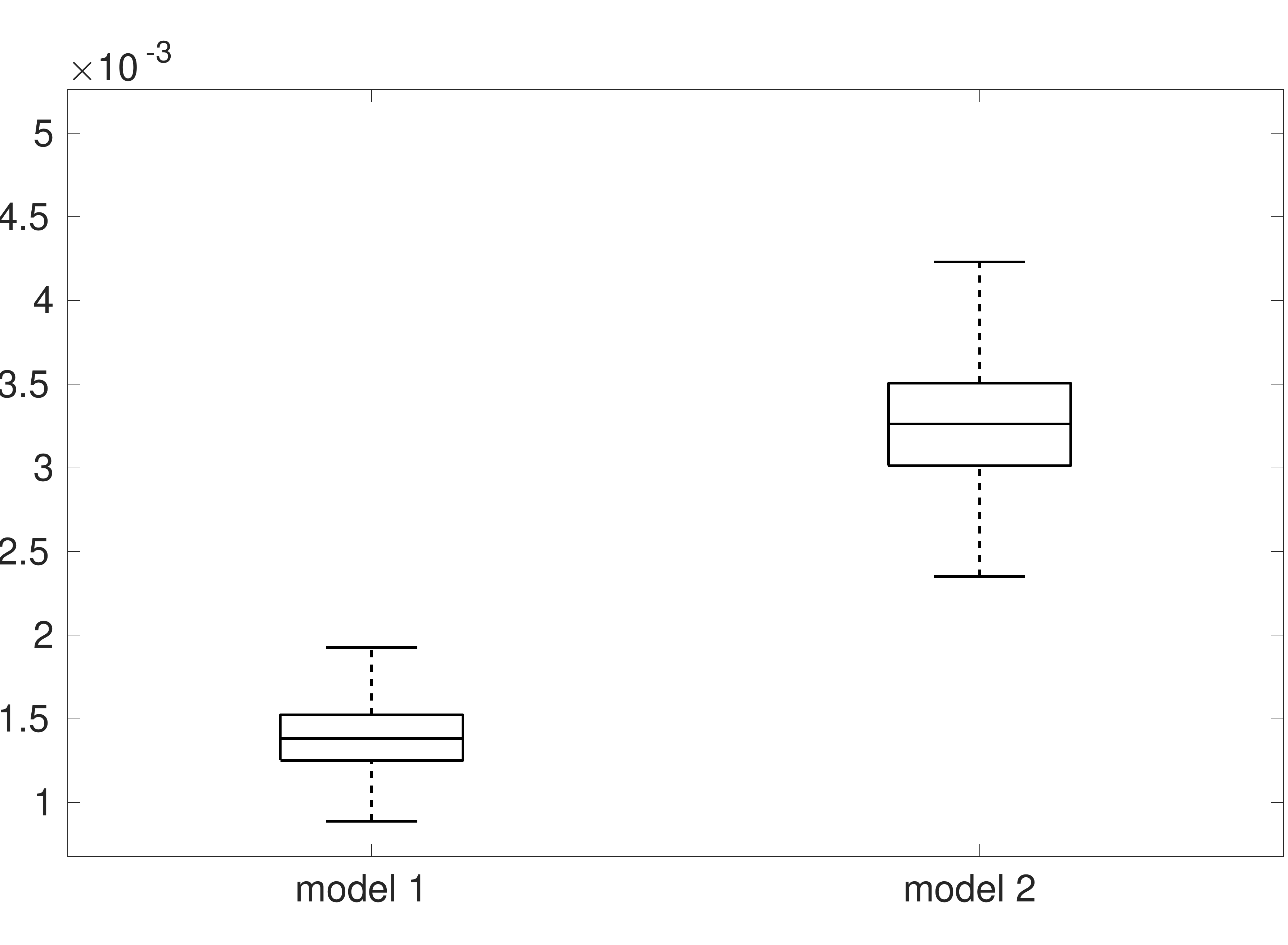}
	\caption{\label{fig:boxplot_WoStLe2017}Box plot without outliers for the $ 0.95 $-quantiles of the  $ n=100 $ absolute model errors for both computer models of the MAFDS. The box plot contains $ B = 500 $ bootstrap $ 0.95 $-quantiles.}
\end{figure}

We apply bootstrap approach from Subsection \ref{se:Wong} to data of the MAFDS and are interested in the $ 0.95 $-quantile of the absolute model error. 
We set $ B = 500 $ and $ n_{l} = 10 $ and define the estimate by \eqref{eq:bootstrap_quantile}.
The resulting medians are shown in Table \ref{ta:results_WoStLe2017}.
\begin{table}[h] 
	\hspace*{-2cm}\centering 
	\begin{tabular}{c|c}
		& MAFDS \\ 
		computer model 1 & $ 0.00138 $ \\
		computer model 2 & $ 0.00326 $ \\ 
	\end{tabular} 
	\caption{\label{ta:results_WoStLe2017} Median of the bootstrap estimates of the $ 0.95 $-quantiles of the $ n=100 $ maximal model errors for both computer models of the MAFDS.} 
\end{table} 
Furthermore Figure \ref{fig:boxplot_WoStLe2017} shows a box plot for both technical systems of the $ B = 500 $ bootstrap estimates of the $ 0.95 $-quantiles.

We can use the method to compare the different computer models and analyse their computer model error.
The predicted computed model error is smaller for computer model 1.
Thus based on this method one can conclude that model 1 predicts the reality better, with an estimated $ 0.95 $-quantile of $ 0.00138 $,

To conclude, the method can be used firstly to compare different computer simulations and secondly to estimate the model error.

\subsection{Comparison}
All methods proved useful in the estimation of the computer model error, although they differ in quality. The AVM of \cite{RoOb2011} is quite simple.
It measures the accuracy of computer models by estimates of their cumulative distribution functions and compares these estimates to estimates of the cumulative distribution function based on experimental data. The difference is measured by the area between these estimates. But the result is hard to interpret.

Using the method of \cite{KeOh2001} to quantify the model error, it is necessary to compute a maximum likelihood estimate. This step is quite unstable since one has to invert matrices, which leads in our examples to an overestimation of the computer model error. In fact in case of the MAFDS the maximal absolute model error of computer model 1 calculated on the $ 100 $ experimental data is $ 0.0023 $ and $ 0.0052 $ for computer model 2, which is $ 70 $ and $ 50 $ times smaller than the estimated $ 0.95 $-quantile of the model error, cf. Table \ref{ta:results_KeOh2001}.

In the applications the method of \cite{WoStLe2017} proved the most accurate. In fact, if we use the $ 100 $ available data points to calculate the empirical $ 0.95 $-quantiles of the absolute model error for both computer models of the MAFDS the resulting values are $ 0.0014 $ for computer model 1 and $ 0.0033 $ for computer model 2. These estimates are quite close to the by the method estimated values.

\section{Quantifying the influence of the computer model error}
\label{se:confidence_estimator}
In the following section, two different methods which analyse the influence of the computer model error on density and quantile estimates are described. The first method aims to estimate a confidence interval for quantiles. The second method aims on estimating a confidence band of the density. With the aid of these methods, it is also possible to compare the influence of the error of computer models.

\subsection{Confidence intervals for quantiles}
\label{se:confidence_interval}
\cite{Koetal2018} proposed a method to construct confidence intervals on quantiles related to experiments with a technical system. The aim of the method is to construct an estimator of the confidence interval of the $ \alpha $-quantile $ q_{Y,\alpha} $ defined by 
\begin{equation}
\label{eq:alpha_quantile}
q_{Y,\alpha} = \min \{ y \in \R \colon G_{Y}(y) \geq \alpha \},
\end{equation}
where
\begin{equation}
\label{eq:cdf_Y}
G_{Y} \colon \R \to [0,1], \quad  G_{Y}(y) = \PROB\{ Y \leq y \},
\end{equation}
i.e. for a given $ \delta \in (0,1) $ they want to construct a (random) interval $ \hat{C}_{n} $ such that
\begin{equation}
\label{eq:confidence_interval}
\PROB \{ q_{Y,\alpha} \in \hat{C}_{n} \} \geq 1- \delta
\end{equation}
holds.

To do this, in a first step a surrogate model $ \hat{m}_{L_{n}} $ of $ m $ as in Subsection \ref{se:surrogate_model} is defined by \eqref{eq:freq_surrogate}.
Then to construct a confidence interval a quantile estimator is constructed. Therefore define an estimator for the cumulative density function of $ \hat{m}_{L_{n}}(X) $ by 
\begin{equation}
\hat{G}_{\hat{m}_{L_{n}}(X),N_{n}}(y) = \frac{1}{N_{n}} \sum_{i=1}^{N_{n}} I_{(-\infty,\hat{m}_{L_{n}}(X_{n+L_{n}+i})]}(y)
\end{equation}
and define the corresponding plug in estimate by
\begin{equation}
\hat{q}_{\hat{m}_{L_{n}}(X),N_{n},\alpha} = \min \{ y \in \R : \hat{G}_{\hat{m}_{L_{n}}(X),N_{n}}(y) \geq \alpha \}
\end{equation}
for $ \alpha \in (0,1) $. 
Next choose some suitable $ \varDelta \delta \in (0,\delta) $, set
\begin{eqnarray*}
	\hat{\beta}_{n} &=& \max_{i=1,\ldots,n} |Y_{i}- \hat{m}_{L_{n}}(X_{i})|,
	\\
	N_{1}(n) &=& N_{n} \cdot \left(\alpha - \sqrt{- \frac{\log( \varDelta \delta /2) }{2N_{n}}} \right),
	\\
	N_{2}(n) &=& N_{n} \cdot \left(\alpha + \sqrt{- \frac{\log( \varDelta \delta /2) }{2N_{n}}} \right),
\end{eqnarray*}
and choose $ \epsilon_{n} \in (0,1) $ such that $ (1-\epsilon_{n})^n < (\delta - \varDelta \delta) $ holds and 
\begin{equation}
\label{eq:eps_gamma}
\epsilon_{n} + \gamma_{n} = \epsilon_{n} + \sqrt{- \frac{\log( \delta - \varDelta \delta - (1-\epsilon_{n})^n ) }{2N_{n}}}
\end{equation}
is minimal.

Finally the $ \delta $ confidence interval for $ q_{Y,\alpha} $ is defined by
\begin{equation}
\hat{C}_{n} = \left[ \hat{q}_{\hat{m}_{L_{n}}(X),N_{n},\frac{N_{1}(n)}{N_{n}}-\epsilon_{n}-\gamma_{n} }-\hat{\beta}_{n}, \hat{q}_{\hat{m}_{L_{n}}(X),N_{n},\frac{N_{2}(n)}{N_{n}}+\epsilon_{n}+\gamma_{n} }+\hat{\beta}_{n}  \right].
\end{equation}

\subsection{Confidence bands for densities}
\label{se:confidence_band}
\cite{KoKr2019} proposed a method to construct confidence bands for the density of $ Y $. The aim of their method is to construct for a finite sample size $ n $ lower and upper bounds $ \hat{g}_{n}^{(lower)} $ and $ \hat{g}_{n}^{(upper)} $ of $ g $ satisfying
\begin{equation}
\label{eq:density_bounds}
\int_{I} \hat{g}_{n}^{(lower)}(x) \, dx \leq \int_{I} g(x) \, dx \leq  \int_{I} \hat{g}_{n}^{(upper)} \, dx
\end{equation}
for all intervals $ I $ with length $ |I|  \geq \kappa_{n} > 0 $, with probability $ 1-\delta $, where $ \delta, \kappa_{n} >0 $ are given values.
To define the estimate the following notation is used:
For $ \beta > 0 $ define
\begin{equation*}
I^{\beta} = \{ x \in \R : [ x-\beta,x+\beta ] \cap I \neq \emptyset \}
\end{equation*}
and
\begin{equation*}
I_{\beta} = \{ x \in \R : [ x-\beta,x+\beta ] \subseteq I \}.
\end{equation*}
The empirical distribution $ \hat{\mu}_{\hat{m}_{L_{n}(X),N_{n}}} $ of $ \hat{m}_{L_{n}}(X) $ is defined by
\begin{equation*}
\hat{\mu}_{\hat{m}_{L_{n}}(X),N_{n}}(A) = \frac{1}{N_{n}} \sum_{i=n+L_{n}+1}^{n+L_{n}+N_{n}} I_{A}(\hat{m}_{L_{n}}(X_{i})) \quad (A \subseteq \R).
\end{equation*}

In a first step a surrogate model $ \hat{m}_{L_{n}} $ of $ m $ is defined as in Subsection \ref{se:surrogate_model} by \eqref{eq:freq_surrogate}.
Then construct a kernel density estimate based on a sample of $ \hat{m}_{L_{n}}(X) $, i.e. the estimator is defined by 
\begin{equation*}
\hat{f}_{\hat{m}_{L_{n}}(X),N_{n},h_{N_{n}}}(y) = \frac{1}{N_{n} \cdot h_{N_{n}}} \cdot \sum_{i = n+L_{n}+1}^{n+L_{n}+N_{n}}
K\left( \frac{y-\hat{m}_{L_{n}}(X_{i})}{h_{N_{n}}} \right),
\end{equation*}
for some bandwidth $ h_{N_{n}} > 0 $ and some kernel $ K \colon \R \to \R $.

Set $ \hat{\beta} = \max_{i=1,\ldots,n} |Y_{i} - \hat{m}_{L_{n}}(X_{i})| $ and choose $ \epsilon_{n} \in (0,1) $ such that $ (1-\epsilon_{n})^n + 2/N_{n}^2 < \delta $ holds for a given $ \delta \in (0,1) $ and that
\begin{equation}
\label{eq:eps_gamma_2}
\gamma_{n} + \epsilon_{n} = \sqrt{ \frac{-\ln(\delta -2/N_{n}^2 - (1-\epsilon)^n)}{2 \cdot N_{n}} } + \epsilon_{n}
\end{equation}
is minimal.
Finally for some given $ \kappa_{n} > 0 $ the upper and lower bounds on the density $ g $ are defined by
\begin{eqnarray}
\hat{g}_{n}^{(upper)}(y) &=& \hat{f}_{\hat{m}_{L_{n}}(X),N_{n},h_{N_{n}}}(y) + \frac{\epsilon_{n}+ \gamma_{n} +\frac{2 \cdot \sqrt{\log N_{n}}}{\sqrt{N_{n}}}}{\kappa_{n}}
\\
&& +
\frac{1}{\kappa_{n}} \cdot \sup_{\substack{J\text{ interval, }y \in J, \\  |J|> \kappa_{n}}}
\Bigg( \hat{\mu}_{\hat{m}_{L_{n}}(X),N_{n}}(J^{\hat{\beta}}) -  \int_{J} \hat{f}_{\hat{m}_{L_{n}}(X),N_{n},h_{N_{n}}}(t) dt
\Bigg)
\nonumber
\end{eqnarray}
and
\begin{eqnarray}
\hat{g}_{n}^{(lower)}(y) &=& \max \Bigg\{ \hat{f}_{\hat{m}_{L_{n}}(X),N_{n},h_{N_{n}}}(y) - \frac{\epsilon_{n}+ \gamma_{n} +\frac{2 \cdot \sqrt{\log N_{n}}}{\sqrt{N_{n}}}}{\kappa_{n}}
\\
&& -
\frac{1}{\kappa_{n}} \cdot \sup_{\substack{J\text{ interval, }y \in J, \\  |J|> \kappa_{n}}}
\Bigg( \int_{J} \hat{f}_{\hat{m}_{L_{n}}(X),N_{n},h_{N_{n}}}(t) dt
- \hat{\mu}_{\hat{m}_{L_{n}}(X),N_{n}}(J_{\hat{\beta}})
\Bigg),0
\Bigg\}.
\nonumber
\end{eqnarray}

\subsection{Application}
In the following, we will illustrate and discuss the above described methods by applying them to data of the technical systems described in Section \ref{se:technical_systems}.
\subsubsection{Confidence intervals for quantiles}
We apply the confidence interval estimator for quantiles from Subsection \ref{se:confidence_interval} to data of the MAFDS and are interested in the $ 95 \% $ (i.e. $ \delta = 0.05 $) confidence interval of the $ 0.95 $-quantile. 
The minimization problem in \eqref{eq:eps_gamma} is approximated by a interior-point method, as implemented by the {\it MATLAB} function {\it fmincon}.

The resulting confidence intervals for the two different computer models are shown in Table \ref{ta:results_Koetal2018}.
\begin{table}[h] 
	\hspace*{-2cm}\centering 
	\begin{tabular}{c|c}
		& MAFDS   \\ 
		computer model 1 & $ [0.0826,0.0890] $   \\
		computer model 2 & $ [0.0786,0.0900] $   \\ 
	\end{tabular} 
	\caption{\label{ta:results_Koetal2018} $ 95 \% $ confidence intervals of the $ 0.95 $-quantile of both computer models of the MAFDS.} 
\end{table} 

In case of the MAFDS, we can use the method to estimate confidence intervals as above which yield more information about the experiment with the technical system. Furthermore it is possible to compare different computer models by analyzing the influence of the computer model error on the estimated confidence intervals.
The predicted confidence interval is narrower for computer model 1 and since we used the same experimental data for both models, the computer model error must be larger for computer model 2.

In case of the piezo-beam, there are only $ 10 $ experimental data values available and thus it is not possible to apply the above method for a reasonable $ \delta $ and $ \alpha $.

To conclude, the method can be used to estimate a confidence interval of some $ \alpha $-quantile. It is only possible to apply it in case a sufficiently large sample is available. In fact for the chosen values of $ \alpha $ and $ \delta $, $ n $ needs to be greater than $ 60 $. For the method to work with $ 10 $ data points and $ \alpha = 0.95 $, one would need to increase $ \delta $ to $ 0.63 $.
Furthermore, the method can be used to analyse the influence of the computer model error on the estimation of the confidence interval. Since the model error directly influences the width of the confidence interval, one can also compare different computer models.

\subsubsection{Confidence bands for densities}
We apply the confidence band estimator from Subsection \ref{se:confidence_band} to data of both technical systems.
The minimization problem in \eqref{eq:eps_gamma_2} is approximated by a interior-point method, as implemented by the {\it MATLAB} function {\it fmincon}.
In both cases we restrict the method to a finite interval. In case of the MAFDS we only consider the interval $ [0.075,0.09] $ and in case if the piezo-beam we consider the interval $ [12,15] $.
We choose $ \kappa_{n} $ as small as possible to get a reasonable confidence band. In case of the MAFDS this means we set $ \kappa_{n} = 0.005 $ and in case of the piezo-beam we set $ \kappa_{n} = 2 $.
The value of $ \delta $ is set to $ 0.05 $.
\cite{KoKr2019} pointed out that the confidence band holds simultaneously for all bandwidths with probability of at least $ 1-\delta $. I.e. we can calculate the confidence band for a multitude of bandwidths and use pointwise the minimal value for the upper band and the maximal value for the lower band.
For computer model 1 of the MAFDS we use $ h_{N_{n}} \in \{0.00005,0.0001,0.0005,0.0025,0.004,0.006\} $, for computer model 2 we use  $ h_{N_{n}} \in \{0.00001,0.0001,0.0005,0.0008,0.001\} $ and for the piezo-beam we use $ h_{N_{n}} \in \{0.005,0.05,0.1,0.2,0.3\} $.
As kernel function $ K \colon \R \to \R $ we use the naive kernel $ K(x) = \frac{1}{2}\cdot I_{[-1,1]}(x) $.

\begin{figure}[h!]
	\centering
	\includegraphics[width=7cm]{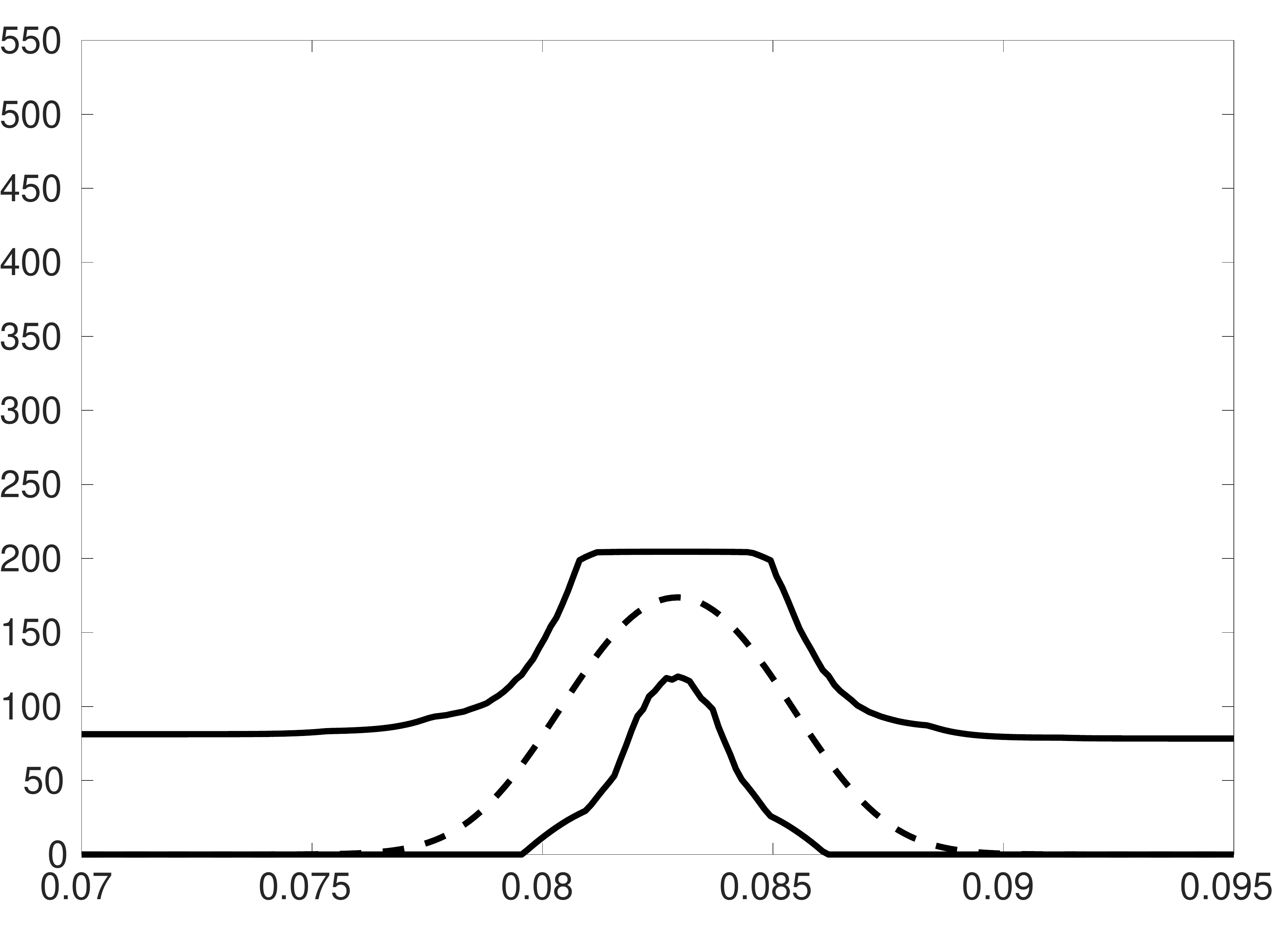}
	\includegraphics[width=7cm]{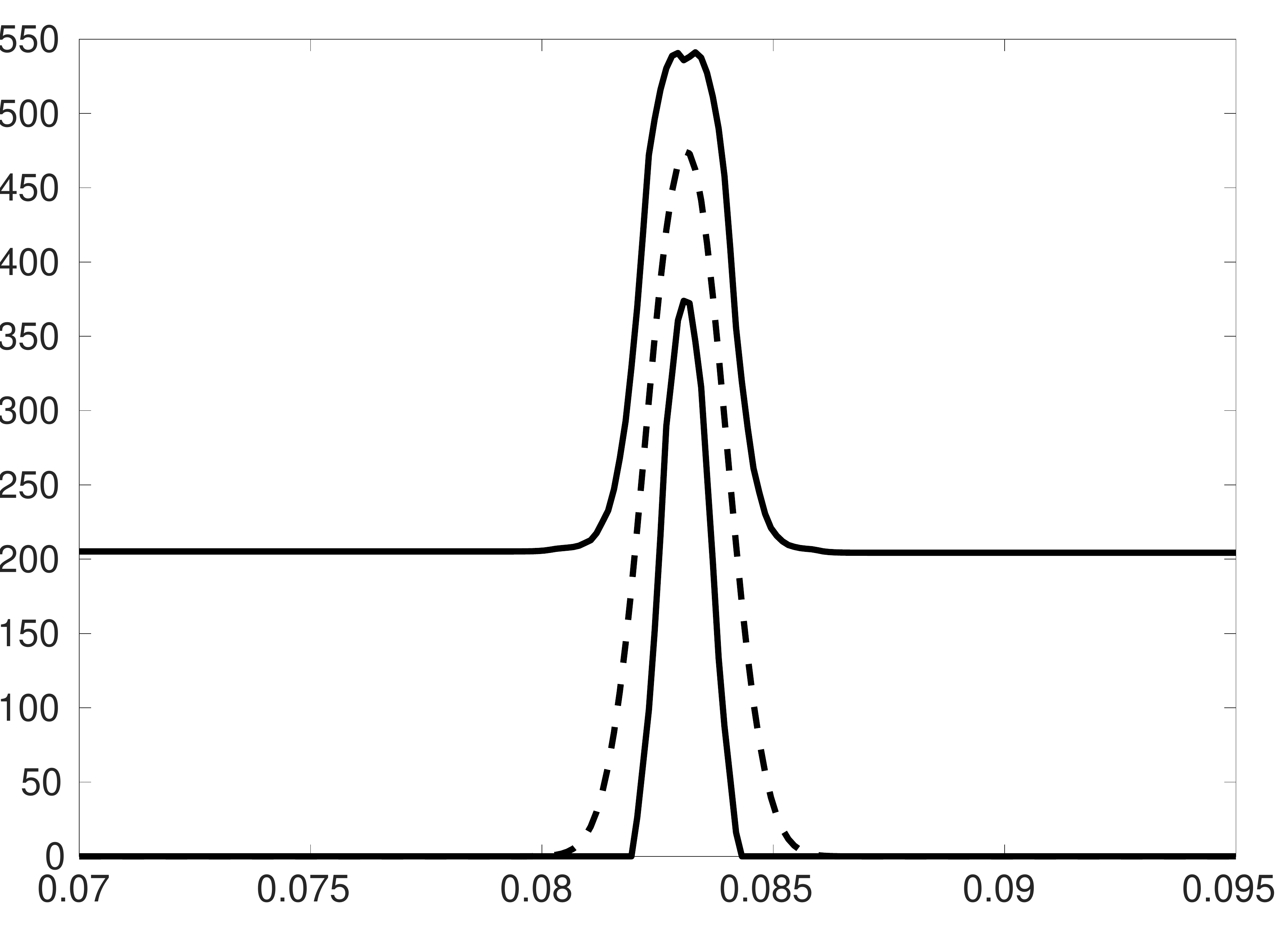}
	\includegraphics[width=7cm]{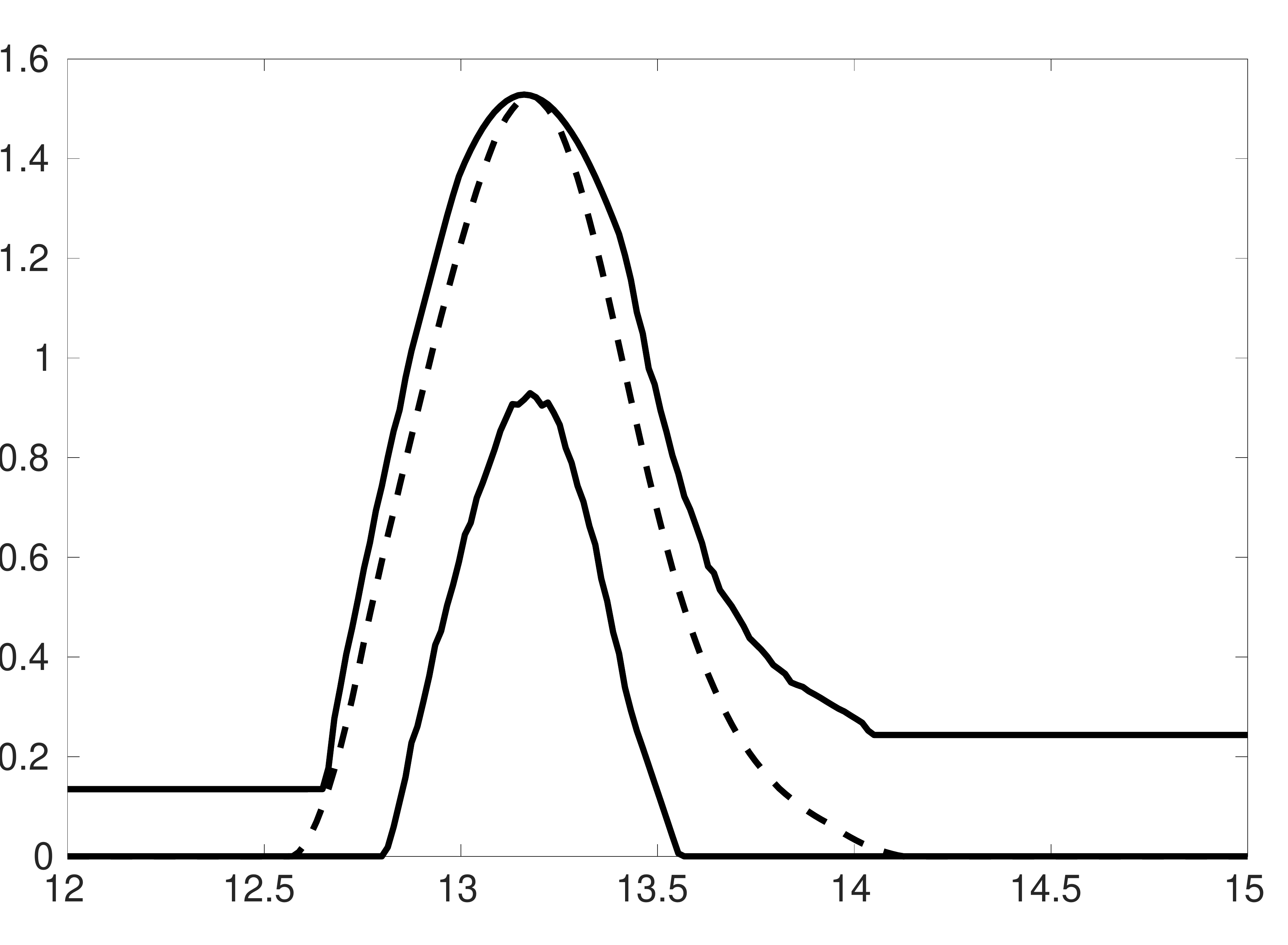}
	\caption{\label{fig:plot_KoKr2019} Upper plots are corresponding to computer model 1 and 2 of the MAFDS, cf. Subsection \ref{se:MAFDS}. Lower plot to piezo-beam, cf. Subsection \ref{se:piezo_beam}. Each plot contains density estimator (dotted line) based on a surrogate model, cf. Subsection \ref{se:density_improved_surrogate} and the upper and lower confidence bands $ \hat{g}_{n}^{(upper)} $ and $ \hat{g}_{n}^{(lower)} $ (upper and lower solid lines). }
\end{figure}

The results are displayed in Figure \ref{fig:plot_KoKr2019}.
In case of the MAFDS, we can use the method to estimate a confidence band of the density of $ Y $ which yields more information about the experiment with the technical system. Furthermore we can compare different computer models and analyse their computer model error.
Here the predicted confidence band is narrower for computer model 1, hence the error of computer model 1 has a smaller influence on the estimation of the density than the error of computer model 2.

In case of the piezo-beam, we can use the method to estimate a confidence band of the density of $ Y $. 
This information can be used in an analysis or the next design stage of the technical system.

To conclude, the method can be used to estimate a confidence band of the density of the outcome of an experiment with a technical system.
Furthermore, the method can be used to analyse the influence of the computer model error on the estimation of the confidence band. Since the model error influences the width of the confidence interval, one can also compare different computer models.
Note that to achieve a reasonable confidence band in case of the piezo-beam, it was necessary to set $ \kappa_{n} $ to a rather large value, so the results here are hard to interpret.

\subsection{Comparison}
The first method can be used to estimate confidence intervals for quantiles, the second to estimate confidence bands for densities. 
By estimating a confidence band for a density one gains more information about the technical system. The problem is, that no real confidence band in a sup norm sense is estimated, i.e. as stated by \eqref{eq:density_bounds}, the estimated confidence band only contains the real density on intervals which are longer than $ \kappa_{n} $ with probability $ 1-\delta $. Thus it is quite hard to interpret the result.

On the other side, the estimated confidence intervals of quantiles do not yield a lot information about the experimental outcome, but it is simpler to analyse the influence of the model error and compare different computer models as illustrated above.

\section{Acknowledgment}
Funded by the Deutsche Forschungsgemeinschaft (DFG, German Research Foundation) - Projektnummer 57157498 - SFB 805.


\bibliographystyle{apalike}
\bibliography{literature}

\newpage
\section*{Supplementary material}

\subsection*{Completion of the square for $ \beta $}

In the sequel we use the following notation
\begin{equation*}
\bar{y} = (y_{1},\ldots,y_{n}), \;
\bar{m} = (m(x_{1}),\ldots,m(x_{n})) \text{ and }
\bar{1} = (1,\ldots,1),
\end{equation*}
where $ \bar{1} $ is also $ n $-dimensional.
In the conditional density 
\begin{eqnarray*}
	&&
	f_{(Y_1,\dots,Y_n)|(\Lambda, \dots, \Omega_d)=(\lambda, \dots, \omega_d)}(y_1, \dots, y_n)
	\\
	&& =  
	\frac{1}{\sqrt{
		(2 \pi)^n \cdot det(\Theta)
	}}
	\cdot
	\exp
	\left(
	- \frac{1}{2}
	\cdot
	(\bar{y}-\bar{m}-\beta\cdot\bar{1})
	\cdot
	\Theta^{-1}
	\cdot
	(\bar{y}-\bar{m}-\beta\cdot\bar{1})^T
	\right)
\end{eqnarray*}
we want to complete the square for $ \beta $. Thus we want to complete the square in the exponent.
Since $ \Theta $ is symmetric and positive definite we have that $ \Theta^{-1} $ is symmetric and positive definite as well. 
Hence
\begin{align*}
& \exp
\left(
- \frac{1}{2}
\cdot
(\bar{y}-\bar{m}-\beta\cdot\bar{1})
\cdot
\Theta^{-1}
\cdot
(\bar{y}-\bar{m}-\beta\cdot\bar{1})^T
\right)
\\
& = 
\exp
\left(
- \frac{1}{2}
\cdot
\left(
\beta^2 \cdot \bar{1} \cdot \Theta^{-1} \cdot \bar{1}^T 
- 2 \cdot \beta \bar{1} \cdot \Theta^{-1} \cdot (\bar{y} -\bar{m})^T
+ (\bar{y} - \bar{m}) \cdot \Theta^{-1} \cdot (\bar{y} - \bar{m})^T
\right)
\right)
\\
& = 
\exp
\left(
- \frac{1}{2}
\cdot
\left(
\bar{1} \cdot \Theta^{-1} \cdot \bar{1}^T
\left(
\beta - \frac{\bar{1} \cdot \Theta^{-1} \cdot (\bar{y} -\bar{m})^T}{\bar{1}\cdot \Theta^{-1} \cdot\bar{1}^T}
\right)^2 
- \frac{\bar{1} \cdot \Theta^{-1} \cdot (\bar{y} -\bar{m})^T}{\bar{1}\cdot \Theta^{-1} \cdot\bar{1}^T}
\right)\right) \cdot
\\
& \qquad \exp \left(
- \frac{1}{2} \cdot
(\bar{y} - \bar{m}) \cdot \Theta^{-1} \cdot (\bar{y} - \bar{m})^T
\right)
\end{align*}
holds. Since $ \Theta^{-1} $ is symmetric and positive definite 
\begin{equation*}
\bar{1} \cdot \Theta^{-1} \cdot \bar{1}^T > 0
\end{equation*}
holds. Thus to maximize the exponent above for $ \beta $ 
\begin{equation*}
\beta = \frac{\bar{1} \cdot \Theta^{-1} \cdot (\bar{y} -\bar{m})^T}{\bar{1}\cdot \Theta^{-1} \cdot\bar{1}^T}
\end{equation*}
needs to hold.

\subsection*{Density}
Since the proposed density for the parameters
\begin{equation*}
p(t) = const \cdot \frac{1}{t}
\end{equation*}
is no real density we propose to use
\begin{equation*}
p(t) = I_{[\epsilon,\exp(1/const) \cdot \epsilon]} \cdot const \cdot \frac{1}{t}
\end{equation*}
instead.

\end{document}